\documentclass[reprint, amsmath, amssymb, aps]{revtex4-1}

\usepackage{graphicx}
\usepackage{dcolumn}
\usepackage{bm}
\usepackage{comment}

\begin{document}

\preprint{APS/123-QED}

\title{Mechanism underlying dynamic scaling properties observed in the contour of spreading epithelial monolayer}

\author{Toshiki Oguma}
\author{Hisako Takigawa-Imamura}%
\author{Takashi Miura}%
\affiliation{%
Department of Anatomy and Cell Biology, Graduate School of Medical Sciences, Kyushu University, 
 Japan
}%

\date{\today}

\begin{abstract}
We found evidence of dynamic scaling in the spreading of MDCK monolayer, which can be characterized by the Hurst exponent $\alpha = 0.86$ and the growth exponent $\beta = 0.73$, and theoretically and experimentally clarified the mechanism that governs the contour shape dynamics. During the spreading of the monolayer, it is known that so- called "leader cells" generate the driving force and lead the other cells. Our time-lapse observations of cell behavior showed that these leader cells appeared at the early stage of the spreading, and formed the monolayer protrusion.
Informed by these observations, we developed a simple mathematical model that included differences in cell motility, cell-cell adhesion, and random cell movement. The model reproduced the quantitative characteristics obtained from the experiment, such as the spreading speed, the distribution of the increment, and the dynamic scaling law.
Analysis of the model equation revealed that the model could reproduce the different scaling law from $\alpha = 0.5, \beta = 0.25$ to $\alpha=0.9, \beta=0.75$, and the exponents $\alpha, \beta$ were determined by the two indices: $\rho t$ and $c$. Based on the analytical result, parameter estimation from the experimental results was achieved. The monolayer on the collagen-coated dishes showed a different scaling law $\alpha = 0.74, \beta=0.68$, suggesting that cell motility increased by $9$ folds. This result was consistent with the assay of the single-cell motility.
Our study demonstrated that the dynamics of the contour of the monolayer were explained by the simple model, and proposed a new mechanism that exhibits the dynamic scaling property.
\end{abstract}

\maketitle


\section{\label{sec:introduction} Introduction}
The shape of mammalian cell colonies varies, depending on the cell type and its environment. In the field of oncology, there is known to be a correlation between shape and malignancy of cancer, and suitable strategies for treatment can be inferred from analyzing the shape of cancer cell colonies \cite{Weinberg_The_2013}. To quantify the shape of these colonies, fractal analysis is often used. It has been reported that a high fractal dimension $\mathcal{D}$ reflects a heterogeneous contour shape, and that fractal dimension and cancer malignancy are likewise correlated \cite{Lennon_Lung_2015}. Cancer cells show higher proliferation rates, higher motilities, and weaker cell-cell adhesion than benign cells \cite{Weinberg_The_2013}. These differences are thought to affect collective cell behavior, and contribute to the roughness of contour shapes in cancer.

Cell movements are promoted by chemical or physical cues such as signal molecules and mechanical forces, in response to which cells modulate their downstream cytoskeleton by altering the subcellular localization of small G proteins such as RhoA, Rac1, and Cdc42 \cite{Chepizhko_Bursts_2016, Mayor_The_2016, Ridley_Cell_2003}.
In collective cell migration, highly motile cells are often observed at the edge of the epithelial monolayer \textit{in vitro}. Known as leader cells, these are characterized by having high Rac1 activity, thick actin filaments, and large cell bodies with spreading lamellipodia. Leader cells can generate a driving force to spread the epithelial tissue, and are often observed at the tips of monolayer protrusions \cite{Omelchenko_Rho_2003, Yamaguchi_Leader_2015, Konen_Image_2017}. However, the relationship between leader cells and the overall contour shape remains controversial. Mark and colleagues suggested that sharp curvature promoted the appearance of the leader cells \cite{Mark_Physical_2010}. On the other hand, several studies have suggested that leader cells are determined by different mechanisms, such as mechanical force \cite{Vishwakarma_Mechanical_2018} and the Dll4-Notch1 pathway \cite{Riahi_Notch1_2015}, then form monolayer protrusions.

It is known that many curves in nature, such as the earth's surface in cross-section and the interface of clouds, form self-affine fractal structures \cite{Voss_Random_1989}. The fractal dimension $\mathcal{D}$ is used to quantify the roughness of the structure. It can be measured using the box-counting method \cite{Mandelbrot_Self_1985, Meakin_Fractals_2011}. For example, the simplest self-affine fractal structure is a Brownian curve, which is the trajectory of a particle undergoing Brownian motion plotted over time, and its fractal dimension is $\mathcal{D}=1.5$. In a self-affine fractal, a similar curve can be obtained by expanding a smaller part of the curve.

If we let $a$ and $b$, the magnification rates for scaling, correspond to coordinate axes in $(1+1)$ dimension, the power exponent $\alpha$, which satisfies $a = b^\alpha$, is called a Hurst exponent. Self-affine fractal structure is characterized by the Hurst exponent $\alpha$, and it is known that $\alpha$ and the fractal dimension $\mathcal{D}$ satisfy $\alpha + \mathcal{D} = 2$ \cite{Mandelbrot_Self_1985,Moreira_On_1994}.

For a growing interface, the local roughness $w(l,t)$ is defined as the standard deviation of the height of the contour, within the closed range $l$. Such a system is said to exhibit dynamic scaling when the following relationships are satisfied \cite{Meakin_Fractals_2011}:
\begin{equation}\label{eq01}
w(l,t) \sim
\left[
\begin{aligned}
 l^{\alpha} &\ \ \text{for } l \ll l_* \\
 t^{\beta}  &\ \ \text{for } l \gg l_*
\end{aligned}
\right.,
\end{equation}
where the length $l^*$ increases as time evolves. The former equation confirms that the interface forms a self-affine fractal structure, and the latter means that the heterogeneity of the whole interface is scaled by the time. The exponent $\alpha$ is the Hurst exponent of the interface, and $\beta$ is called the growth exponent.

Dynamic scaling has been identified in a wide range of phenomena, including bacteria colonies on agar \cite{Wakita_Self_1997, Santalla_Nonuniversality_2018}, propagation of slow combustion of paper \cite{Myllys_Kinetic_2001}, and growing interfaces in a turbulent liquid crystal \cite{Takeuchi_Evidence_2012}. In collective migration of vertebrate cells, the contours of cell colonies of HeLa and Vero cell lines were also found to exhibit dynamic scaling, with $\alpha = 0.50$ and $\beta = 0.32$ \cite{Huergo_Growth_2012, Muzzio_Influence_2014}. 
Some partial differential equations that show the properties of dynamic scaling are well-known, including the Edwards-Wilkinson (EW) equation, the Kardar-Parisi-Zhang (KPZ) equation, and the Kuramoto-Sivashinsky (KS) equation.
The EW equation is a stochastic partial differential equation with diffusion and spatiotemporally independent noise terms \cite{Edwards_The_1982}, the KPZ equation is similar to the EW equation but includes a non-linear term \cite{Kardar_Dynamic_1986}, and the Kuramoto-Sivashinsky equation is a partial differential equation with the diffusion, non-linear, and spatially fourth derivative terms \cite{Roy_The_2019}.
Intriguingly, both the KPZ and KS equations are known to have the same Hurst and growth exponents, $\alpha = 1/2$ and $\beta = 1/3$. The sharing of similar scaling exponents between different phenomena and solutions is a property known as universality. In particular, the universality characterized by $\alpha = 1/2$ and $\beta = 1/3$ is called KPZ universality \cite{Kardar_Dynamic_1986, Takeuchi_An_2018}.

While various experimental systems that show dynamic scaling have been reported, to our knowledge, the underlying mechanisms that generate dynamic scaling laws are rarely considered \cite{Santalla_Nonuniversality_2018}.
In the current study, we aimed to uncover the mechanisms that lead to dynamic scaling properties in the contour of the epithelial monolayer, through experimental observations, numerical simulation, and analysis. In experiments, we observed the spreading of the Madin-Darby canine kidney (MDCK) cell monolayer and found that its time evolution followed a dynamic scaling law that was distinct from the KPZ universality.
Our observations and mathematical model showed that the emergence of leader cells and cell-cell adhesion both play critical roles in the dynamics of the contour. From the analytical consideration, we show that $\alpha$ and $\beta$ are determined by the index $c$, which reflects the relative intensity of the random movement on differences of cell motility.


\section{\label{sec:mm}Materials and Methods}
\subsection{MDCK cell culture}
MDCK II cells were cultured using Eagle's minimal essential medium (MEM; Nacalai tesque) containing 10 \% fetal bovine serum (FBS) and 100 U/ml penicillin-streptomycin (Nacalai tesque), and maintained in a 5 \% $\text{CO}_2$ controlled atmosphere at 37 \(^\circ\)C. 

\subsection{Fabrication of PDMS sheets}
Polydimethylsiloxane (PDMS) sheets, used to create a cell-free regions, were prepared in the following manner: A well-mixed PDMS (Sylgard 184, Toray) precursor solution, with a 10:1 ratio of prepolymer to curing agent, was poured onto a flat polystyrene plate to a thickness of 1 mm, then cured in a drying oven at 80 \(^\circ\)C for 2 hr. After curing, 8-mm PDMS disks with a 3- or 4-mm holes were created using biopsy punches (Maruho). 

\subsection{Cell patterning}
The PDMS sheets were placed on the surface of a 27 mm glass bottom dish (IWAKI), and the holes in the sheets were filled with MEM (without air bubbles), along with 2 ml of MDCK cells suspended in medium ($2.5 \times 10^5$ cells/ml). Samples were incubated at 37 \(^\circ\)C and 5 \% $\text{CO}_2$ for 48-72 hr, until cells reached confluence within the holes, then the PDMS sheets were gently removed and washed with MEM twice. The medium was changed with fresh medium containing CellTracker Green CMFDA Dye ($5 \mu M$; Invitrogen) and Hoechst 33342($1 \mu g/ml$; Dojindo). After 4 hr incubation, the time-lapse observation was performed. The collagen-coated dishes were prepared with 27 mm glass bottom dishes and type I-c collagen solutions ($100 \mu g/ml$; Nitta Gelatine).

\subsection{Time-lapse microscopy}
The time-lapse observations were performed using a Nikon A1 confocal microscope with a 10$\times$ or 20$\times$ objective lens. The cells were maintained in a 5 \% $\text{CO}_2$ controlled atmosphere at 37 \(^\circ\)C. Images were acquired every 20 min (for Fig. \ref{fig:3}) or 30 min (for Fig. \ref{fig:2}, \ref{fig:6}, S5) until 18 hr had passed after the PDMS sheet removal. 

\subsection{Measurement and quantification of cell spreading}
The green fluorescent microscopy images were numerically converted and analyzed quantitatively using ImageJ (NIH), Python, and Mathematica (Wolfram). The images were binarized to capture the shape of the monolayer, with the threshold values for binarization determined and set manually. We defined the centroid as the center of gravity of the cell monolayer in the initial image. The distance $D(\theta)$ from the centroid to the contour was measured along 2000 directions, with constant intervals. Then, the mean front distance was calculated as $\langle D(\theta) \rangle(t) = \sum_{\theta} D(\theta) / 2000$. The Hurst exponents were calculated as the slopes of the fitted lines in the log-log plot of $w(l,t)$ and $l$ within the range of $l \in [2\Delta x, 10 \Delta x]$. Here, $\Delta x$ was defined by $2 \pi \langle D(\theta) \rangle /2000$. The growth exponents were calculated as the slope of the fitted lines in the log-log plot of $w(l_\text{max},t)$ and $t$. The values shown in Figs. \ref{fig:2}(b,c,e), \ref{fig:6}(a), and S5(c) are averages of the experimentally obtained values.

\subsection{Cell tracking}
We labeled cell nuclei with Hoechst33342, and images were obtained from 1 hr to 18 hr after PDMS removal and analyzed with ImageJ (NIH), using the TrackMate plugin \cite{Tinevez_TrackMate_2017} to manually track the centers of cell nuclei. We considered cells whose nuclei were located within $30 \mu m$ of the contour to be at the edge of the monolayer. The contours were determined from the brightfield images, and manually traced with segmented lines.

\subsection{Cell staining}
MDCK cells were washed with phosphate-buffered saline (PBS), fixed with 4\% paraformaldehyde (PFA) for 10 min, and permeabilized in 0.1 \% Triton X-100/PBS for 10 min. After washing, cells were stained with Alexa Fluor 555 Phalloidin (1:40; Invitrogen) and Hoechst33342 (1:2000; Dojindo). Cells were incubated for 30 min at room temperature before observation. Fluorescent and phase-contrast images were acquired using a BZ-X810 fluorescence microscope (Keyence) with 20$\times$ objective lens.

\subsection{Numerical simulation}
The numerical simulations were performed using Mathematica and Julia, and we used the explicit Euler scheme for calculating the time evolution. The code used for this study is available from the corresponding authors upon request.

\subsection{Single cell tracking assay}
MDCK cells were seeded onto the 27 mm glass bottom dish ($2.5 \times 10^4$ cells per ml). The cells were incubated at 37 \(^\circ\)C and 5 \% $\text{CO}_2$ for 24 hr until the cells adhered to the bottom of the dish. Hoechst33342 (1:2000) was added to the dish and incubated for 30 min. Images were acquired using a Nikon A1 confocal microscope at 5 min intervals over a span of 2 hr. The centers of cell nuclei were manually tracked using the TrackMate plugin \cite{Tinevez_TrackMate_2017} in ImageJ, and statistical analysis was performed in Mathematica, using Student's t-tests to compare experimental groups.

\section{\label{sec:results}Results}
\subsection{\label{sec:results1}Epithelial monolayer spreading}
\begin{figure}[hbt]
\includegraphics[width=8.6cm]{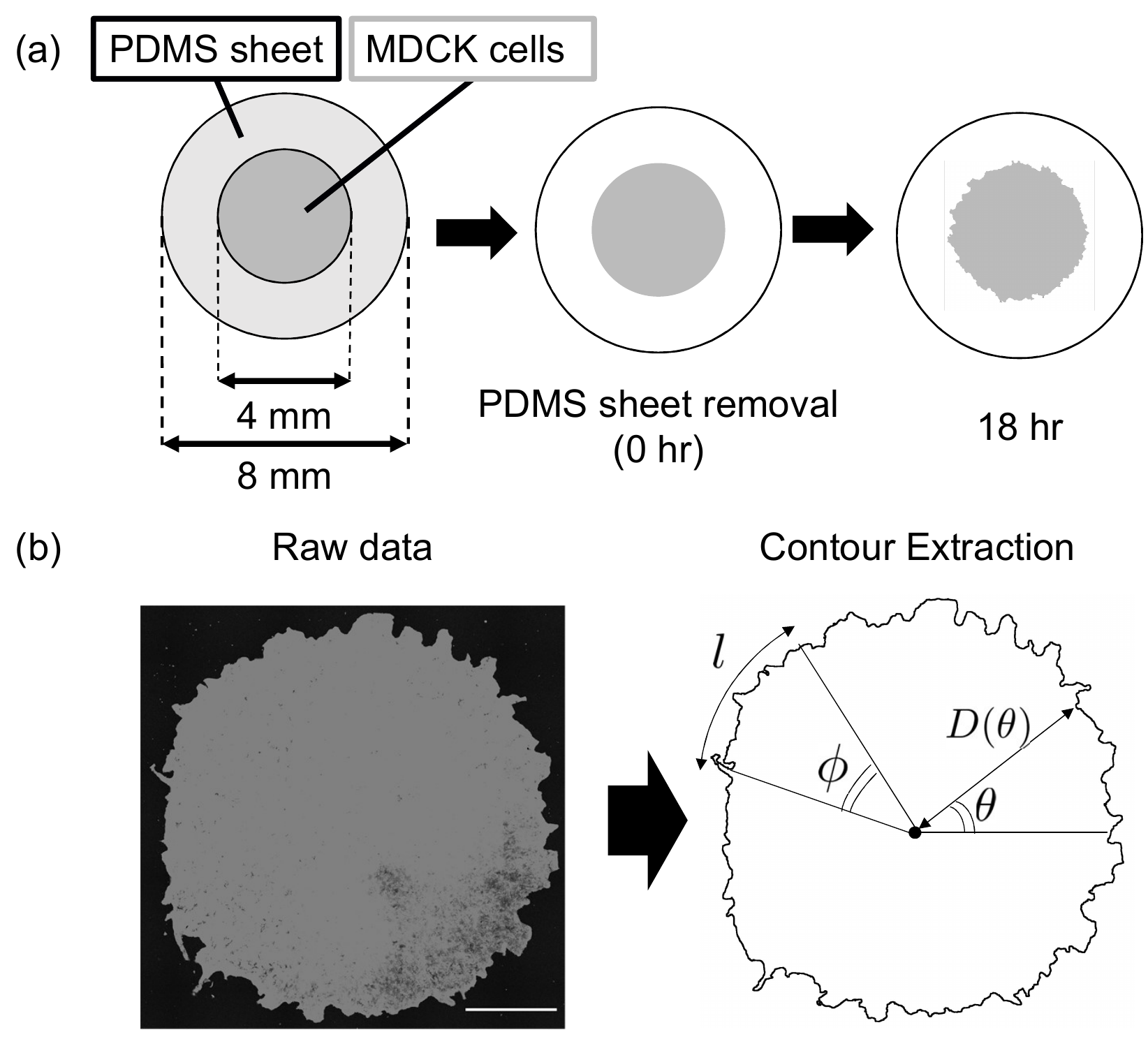}
\caption{\label{fig:1} 
Overview of the experimental method and measurement procedures. 
(a) Observation of MDCK migration. MDCK cells were seeded in the region confined by the PDMS sheet. After cells reached confluence, the PDMS sheet was removed and time-lapse observations were performed. (b) Measurement of the contour shape. The MDCK monolayer was visualized by Cell Tracker Green, and the microscopy images were transformed into binarized images. The distances $D(\theta)$ from the center to the edge of the monolayer were measured. The orientation of the points along the contour was defined as $\theta$ $(0<\theta<2\pi)$. The angle $\phi = l / \langle D \rangle(t)$ was defined from the unit of measurement $l$. Scalebar = 1000 $\mu m$.}
\end{figure}

To investigate how the epithelial monolayer spread, MDCK cells were cultured in the closed circular area confined within the PDMS sheet. Time-lapse observations were performed after removal of the PDMS sheet boundary (Fig. \ref{fig:1}(a), Supplemental Movie 1). 

\begin{figure}[hbt]
\includegraphics[width=8.6cm]{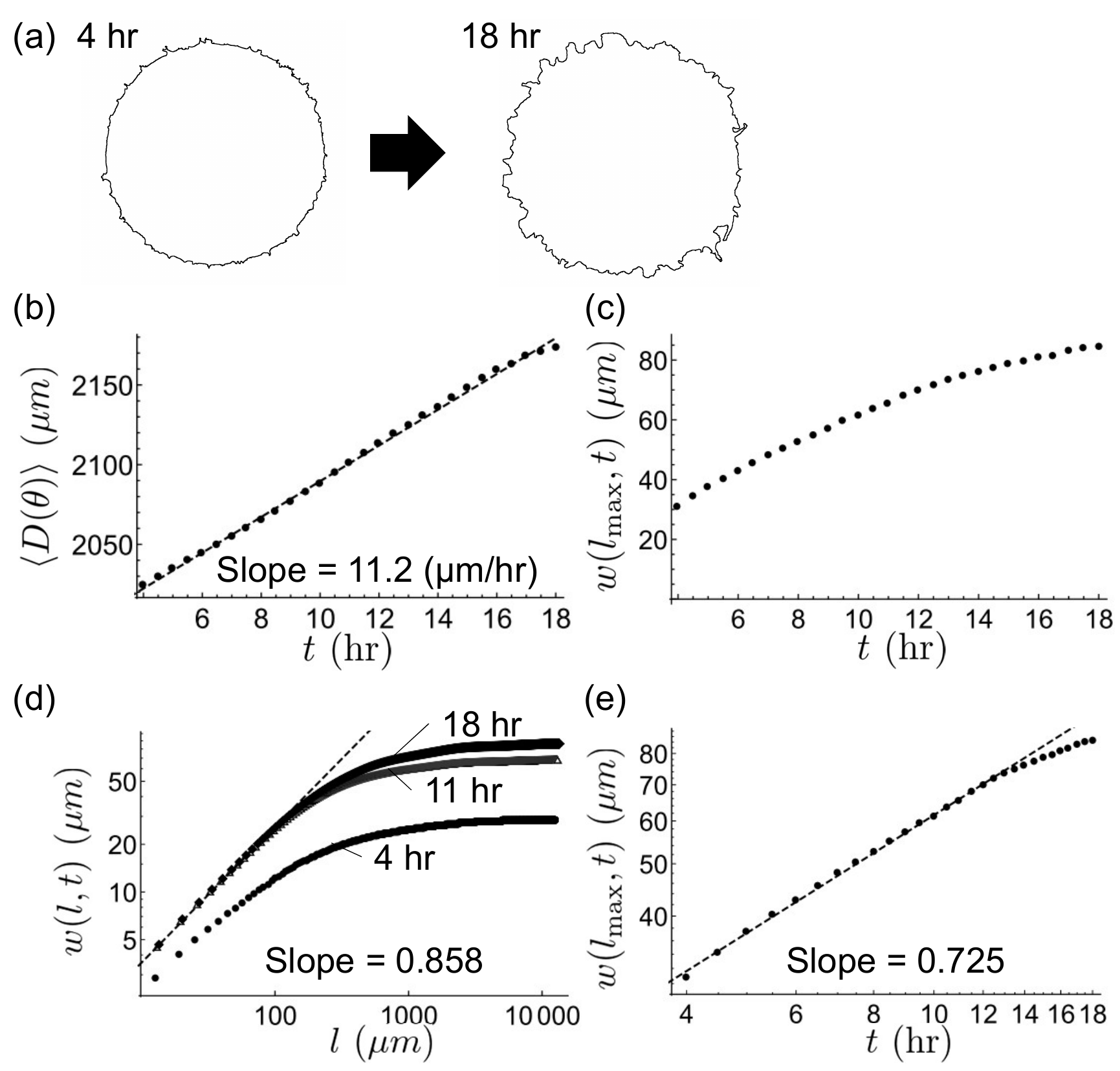}
\caption{\label{fig:2} 
Dynamic scaling law in the spreading of MDCK monolayer, obtained from experimental observations.
(a) Contour of the MDCK monolayer, as it evolved from the initial smooth shape to the rough shape. (b) Time evolution of the average values of $D(\theta)$. Dots indicate the measured values, with the fitted line shown as dashed (n=6). (c) Time evolution of $w(l_\text{max},t)$, the standard deviation of $D(\theta)$ (n=6). (d) Measurement of Hurst exponent. The plot shows the experimental results at different time points, and the black dashed line is the fitted line for $2 \Delta x \leq l \leq 10 \Delta x (\mu m)$ at 18 hr. From this, the Hurst exponent was calculated as: $\alpha = 0.858$. (e) Log-log plot of $w(l_\text{max},t)$ and $t$. Dots represent the measured values, and the black dashed line is the fitted line (n=6). The growth exponent was calculated as: $\beta = 0.725$ }

\end{figure}

While the contour of the epithelial monolayer was initially smooth and round, our observations revealed that it became rougher and more uneven over time, as cells migrated to the cell-free area. The epithelial cells kept contact each other (Supplemental Movie 1). To quantify this, we measured the distance from the center to the edge of the monolayer, $D(\theta)$, where $\theta$ indicates the orientation of measurement points along the contour $(0<\theta<2\pi)$ (Fig. \ref{fig:1}(b)). We chose 2000 points for measuring $D(\theta)$, to accurately reproduce the contour boundary during migration. The initial circumference (after the PDMS removal) was $1.26 \times 10^4\ \mu m$. We found that the average $\langle D(\theta) \rangle$ increased linearly, at a rate of $11.2\  \mu m /hr$, and the standard deviation of $D ( \theta)$ also increased (Fig. \ref{fig:2}(b) and \ref{fig:2}(c)). These results suggested that the epithelial monolayer spread at a constant speed with increasing heterogeneity of the contour.

We characterized the local roughness $w(l,t)$ of the circular geometry as follows:
\begin{eqnarray}\label{eq02}
w(l, t) &=& 
\left\langle
  \sqrt{ 
    \left\langle
      \left[D(\theta, t)-\langle D\rangle_{\phi} \right]^{2} 
    \right\rangle_{\phi}
  }
\right\rangle \\
\phi &=& \frac{l}{\langle D\rangle} \nonumber,
\end{eqnarray}
where $t$ (hr) is the time after PDMS removal, and $\langle \cdots \rangle_{\phi}$ denotes the average value in the area $\phi$. The local roughness $w(l,t)$ is the standard deviation of the distance $D(\theta)$ in the range of $\phi$.

To examine whether the power law in \eqref{eq01} holds for the cell spreading, log-log plots were used to assess the relationship between local roughness $w(l,t)$ and $l$ at different time points (Fig. \ref{fig:2}(d)). Within the range of small $l$, $\ln{w(l,t)}$ linearly increased along with increasing $\ln{l}$, indicating that the power law $w(l,t) \sim l^{\alpha}$ holds, which suggests the contour of the epithelial monolayer is a self-affine fractal structure. We also found the range of $l$ that showed a linear relationship to expand over time, which is consistent with the well-known observation that $l^*$ increases with time under the dynamic scaling law \cite{Meakin_Fractals_2011, Takeuchi_An_2018}.
We determined the Hurst exponent to be $\alpha = 0.858$ for $t=18$. For large $l$, the value of $w(l,t)$ approaches $w(l_\text{max},t)$, where $l_\text{max} = 2\pi \langle D \rangle$. As shown in Fig. \ref{fig:2}(e), $\ln w(l_\text{max},t)$ and $\ln{t}$ show a linear relationship, indicating that the power law $w \sim t^{\beta}$ holds. The growth exponent was $\beta = 0.725$. Thus, our observations revealed that the time development of the MDCK monolayer contour satisfied the dynamic scaling law.
We repeatedly measured the exponents in different monolayers, and obtained values of $\alpha = 0.853 \pm 0.013$ and $\beta = 0.757 \pm 0.065$, which were not much different from the data shown in Fig. \ref{fig:2}.

\subsection{\label{sec:results2}Cell behavior at the edge}
\begin{figure*}[htb]
\includegraphics[width=17.8cm]{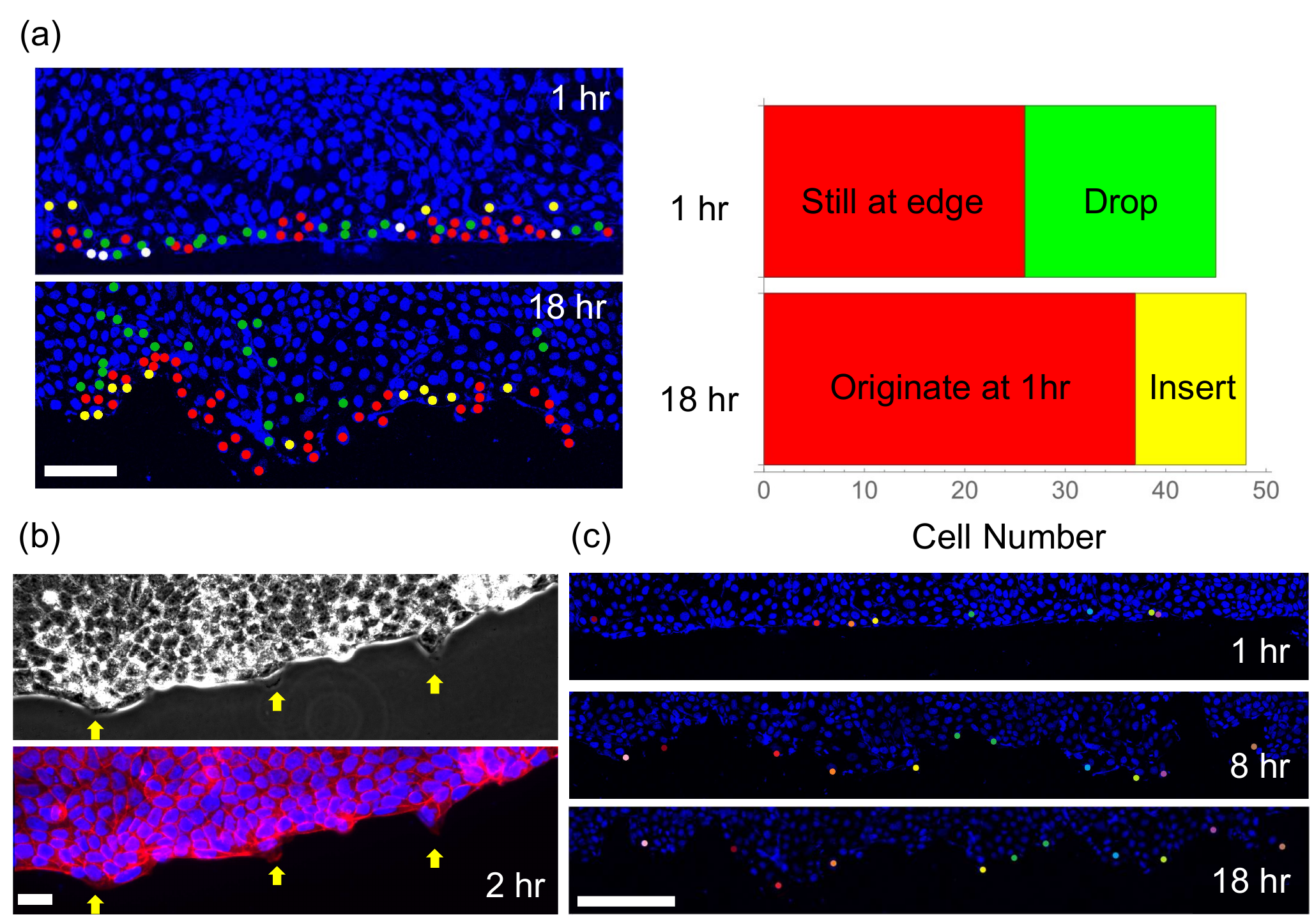}
\caption{\label{fig:3} (Color) Behavior of the cells at the edge of the monolayer. 
(a) Cell tracking at the monolayer edge. Nuclei were visualized with Hoechst33342 (left panels, blue circles). Cells that remained near the edge from $t = 1$ to $t = 18$ are indicated by red circles. Those cells that dropped and intercalated from the edge by $t = 18$ are indicated by green and yellow circles, respectively. Among $45$ cells initially located at the edge, $19$ cells dropped, while $11$ cells divided (right panel). $11$ cells newly intercalated to the edge, therefore, 77 \% of the edge cells at $t = 18$ are originated from the cells at the edge at $t = 1$. Scale bar = $100 \mu m$. (b) Snapshot of the edge at $t = 2$. Large cells that stretch their lamellipodia were observed (indicated with yellow arrows). Upper panel is a phase-contrast image, and the lower panel is a fluorescent image; the nuclei and F-actin are represented in blue and red, respectively. (c) Behavior of leader cells during collective cell migration. Circles indicate leader cells, which are distinguished by their locations and large sizes. 
Scale bar = $200 \mu m$.}
\end{figure*}

To understand the dynamics of the evolving contour shape as the monolayer spreads, we observed and quantified the behavior of cells near the monolayer edge. Cell nuclei were visualized and tracked from $t=1$ to $t=18$ (Fig. \ref{fig:3}(a), Supplemental Movie 2). Among cells initially located in the edge region, 58\% of these remained near the edge the entire observation time, while 42\% migrated towards the monolayer center. These internalizations were observed to result from the merging of monolayer protrusions. The decrease in the number of cells around the edge was compensated for by proliferation and the intercalation of internal cells. We reversely tracked cells near the edge at $t=18$, and found that 77 \% of these were also near the edge at $t=1$. In addition, the kymograph of the edge cell migration showed that movement towards the initially cell-free region first started among edge cells, then the inner cells followed (Fig. S1). These results suggested that the dynamics of the contour shape were primarily driven by the movements of cells at or near the edge.

Since leader cells are known to play an important role in the formation of monolayer protrusion, we next focused on their behavior and dynamics. Leader cells are characterized by their large lamellipodia, large cell bodies, and high motility. Cells with these characteristic shapes were observed as early as $t=2$ (Fig. \ref{fig:3}(b)). As shown in Fig. \ref{fig:3}(c), we performed reverse tracking of the leader cells from $t=18$ to $t=1$. Leader cells were clearly distinguished by their locations and large cell bodies at $t=18$. They were consistently at the edge, formed the monolayer protrusions, kept located at the tip, and had higher velocities than other cells (Fig. S2). The fact that leader cells kept located at the tip of the protrusion reflected its spontaneous high motility, and suggested that the high motilities of the leader cells were maintained throughout the observation period. These results indicate that leader cells emerged at an early stage of the migration at the edge, and that their properties did not change.

\subsection{\label{sec:results3}Mathematical model and numerical simulation}
\begin{figure}[hbt]
\includegraphics[width=8.6cm]{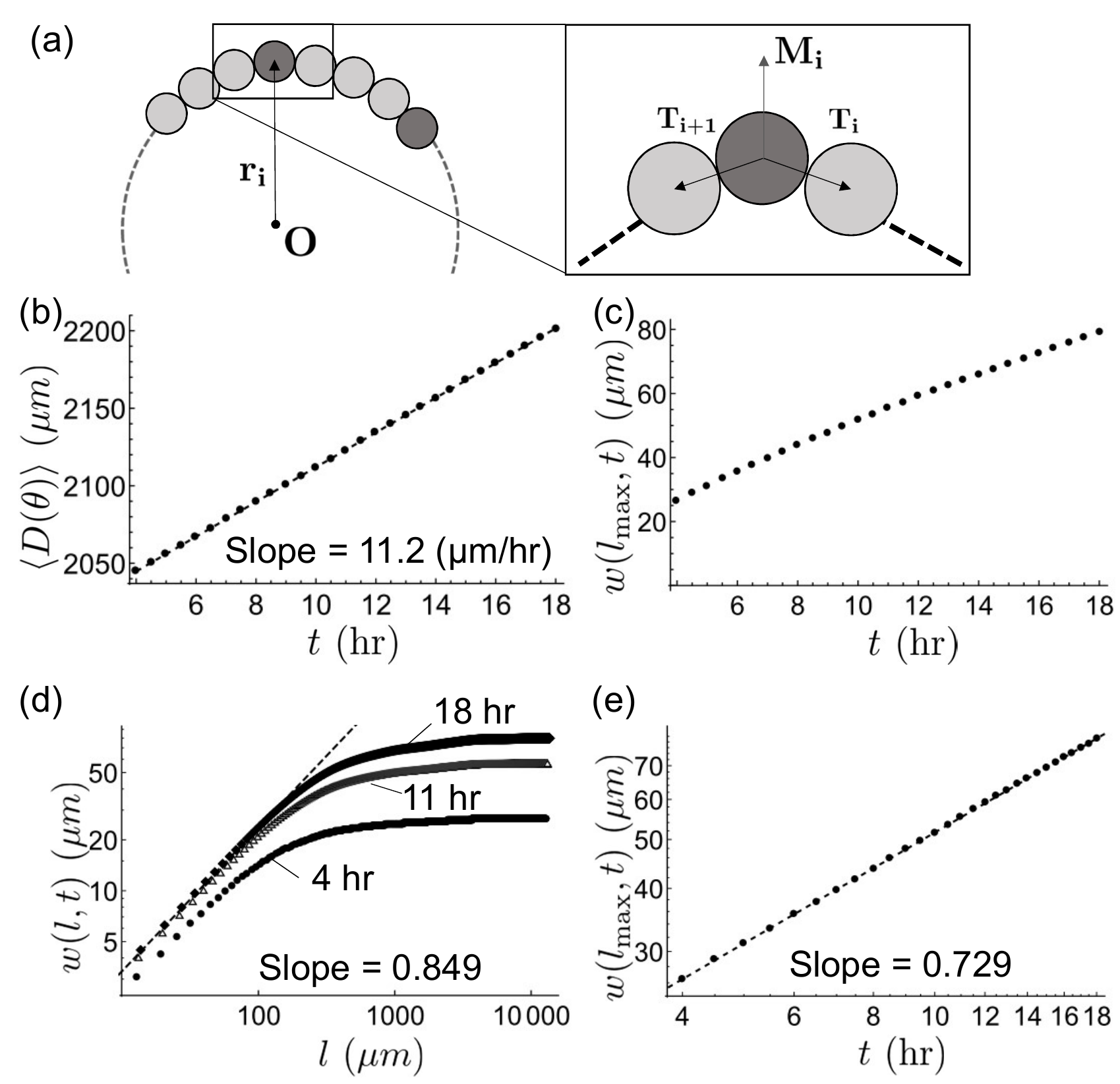}
\caption{\label{fig:4} Numerical simulation of the monolayer spreading. 
(a) Diagram of the mathematical model. Circles represent the cells at the edge, with the dark circles representing leader cells, and the light circles follower cells. The coordinates of the cell center are denoted by ${\bf {r_{\it i}}}$. Cells at the edge have different types of motility ${\bf {M_{\it i}}}$, and interact with neighboring cells through cell-cell adhesion ${\bf {T_{\it i}}}$
(b-e) Results from the numerical model corresponding to Fig. \ref{fig:2}(b) to \ref{fig:2}(e). (b) Time evolution of the average of the distance $D(\theta)$.  (c) Time evolution of the standard deviation of $D(\theta)$ (global roughness). (d) Log-log plot of the local roughness $w(l,t)$ against $l$. (e) Log-log plot of the global roughness $w(l_\text{max},t)$ against $t$. Parameters: $v_l = 94 \ [\mu m\ h^{-1}]$, $v_f = 2 \ [\mu m \ h^{-1}]$, $p_l =0.1$, $\rho = 6.67 [h^{-1}]$, and $\sigma=3 \ [\mu m \ h^{-1/2}]$.}
\end{figure}

Informed by the experimental results, we modeled the dynamics of the contour of the MDCK monolayer. We assumed that the contour dynamics arise from the movements of cells at the edge, and these cells have different motility. The model also assumes, based on a known property of epithelial cells, that cells interact through intercellular adhesion (Fig. \ref{fig:4}(a)). We described the dynamics of cells at the edge by means of temporally-continuous and spatially-discrete differential equations as follows:
\begin{equation} \label{eq03}
  \frac{d}{dt} {\bf {r_{\it i}}} (t) = {\bf {M_{\it i}}} + ({\bf {T_{\it i}}}-{\bf {T_{\it {i+1}}}}) + \sigma \eta_{i}(t) {\bf {r_{\it i}}}/|{\bf {r_{\it i}}}|,
\end{equation}
where ${\bf {r_{\it i}}}(t)$ represents the coordinates of the cell $i$ at time $t$, ${\bf {M_{\it i}}}$ is active, directional movement, and ${\bf {T_{\it i}}}$ describes passive movement due to tension from cell-cell adhesion. The third term on the right-hand side represents random movement, with the constant $\sigma$ indicating the intensity of noise, while $\eta_{i}(t)$ is spatially and temporally independent Gaussian white noise. At the cellular scale, the inertial force may be ignored.

The directional motility term ${\bf {M_{\it i}}}$ was given by
\begin{equation}\label{eq04}
{\bf {M_{\it i}}} = 
\left[
  \begin{aligned}
   v_l \ {\bf {r_{\it i}}}/|{\bf {r_{\it i}}}| &\quad& \text{(if the cell $i$ is a leader cell)} \\
   v_f \ {\bf {r_{\it i}}}/|{\bf {r_{\it i}}}| &\quad& \text{(otherwise)},
  \end{aligned}
\right. 
\end{equation}
where $v_l$ and $v_f$ are the velocity of the leader and follower cells, respectively ($v_l > v_f > 0$). The ratio of leader cells to all edge cells is $p_l$, and the distribution of leader cells was randomly determined.

Assuming that intercellular tension ${\bf {T_{\it i}}}$ linearly increases with the intercellular distance, ${\bf {T_{\it i}}}$ was given by
\begin{equation}\label{eq05}
{\bf {T_{\it i}}} = \rho ({\bf {r}_{\it i}} - {\bf {r}_{\it i-1}}),
\end{equation}
where $\rho$ is the tension coefficient. 

We set the initial shape of the monolayer as a circle with a radius of $2000 \ \mu m$. The total number of cells at the edge was set to $N=2000$, aligned with a regular interval. For simplicity, the cell number change and rearrangement at the edge were omitted. The numerical results are shown in Fig. \ref{fig:4} and Supplemental Movie 3. 
The parameter $p_l$ was estimated from the actual distribution of cells displaying large lamellipodia (Fig. \ref{fig:3}(b)), and $v_l$ and $v_f$ were set such that the growth speed of the average diameter was consistent with the experimentally observed values.

We found that the model \eqref{eq03} described and captured the process of cell sheet expansion well, as it was able to reproduce many of the properties observed in our experiments, as shown in Fig. \ref{fig:4}(b) to \ref{fig:4}(e). The time evolution of $\langle D (\theta) \rangle$ was linear, with a slope of $11.2 \ \mu m/h$. The values of $w (l_\text{max}, t)$ shown in Fig. \ref{fig:4}(c) were similar to those in Fig. \ref{fig:2}(c). The log-log plot of $w(l,t)$ against $l$ showed a linear relationship for small $l$ (Fig. \ref{fig:4}(d)), and the Hurst exponent was calculated as $\alpha = 0.849$. The log-log plot of $w (l_\text{max}, t)$ against $t$ also showed a linear relationship (Fig. \ref{fig:4}(e)), and the growth exponent was calculated as $\beta = 0.729$. In addition, the time evolution of $\text{Max}[D(\theta)]$, $\text{Min}[D(\theta)]$ and the distribution of the increment for $\theta$ were also similar to the experimental results (Fig. S3). Taken together, these results show that the model \eqref{eq03} explained the dynamic scaling law seen in the contour of the MDCK monolayer.

\subsection{\label{sec:results4}Analysis of mathematical model}
In this section, we show that the Hurst and growth exponent were analytically estimated, then evaluate the effects of the cell-cell adhesion, the difference of cell motility, and the noise intensity in this system.

Assuming that $N$ is sufficiently large, it can be regarded that the tension affects only the radial direction, thus the model was simplified to a one-dimensional flat model with periodic boundary as follows:
\begin{equation} \label{eq06}
\frac{d}{d t} h_x(t)= f_x + \rho (h_{x+1}(t)+h_{x-1}(t)-2h_{x}(t) ) + \sigma \eta_x(t).
\end{equation}
Here $h_x(t)$ is the distance from the center to the cell $x$ (for simplicity, set $h_x(0) = 0$), and $f_x$ is the motility of the cell $x$ that takes $v_l$ or $v_f$. 
The cell with $f_x =v_l$ was selected randomly with the ratio of $p_l$. The numerical calculation of the flat model \eqref{eq06} reproduced the characteristic dynamics with the dynamic scaling law generated by the circular model \eqref{eq03} (Fig. S4).

The exponents in the model \eqref{eq06} take different values depending on the parameters. When $v_l = v_f$, the model \eqref{eq06} is essentially the EW model, which includes the Gaussian white noise and the diffusion term. It is known that the EW model shows dynamic scaling $\alpha = 0.50$ and $\beta=0.25$ \cite{Meakin_Fractals_2011, Edwards_The_1982}. On the other hand, if $\sigma = 0$, the model \eqref{eq06} is regarded as the model with temporally fixed noise and diffusion, which shows the dynamic scaling law: $\alpha = 0.9$ and $\beta = 0.75$. Tentatively, we called these dynamics the fixed noise model.

Considering the discrete Fourier expansions of $h_x(t)$, $f_x$, and $\eta_x(t)$, the Fourier coefficients are denoted by $\hat{h_k}(t)$, $\hat{f_k}$, and $\hat{\eta_k}(t)$, respectively. The random variable $\hat{f_k}$ follows Gaussian distribution denoted as $(v_l-v_f) \sqrt{p_l(1-p_l)} \hat{\mathcal{N}}(0,1)$ (See the supplemental text A). The complex Gaussian white noise $\hat{\eta_k}(t)$ satisfies $\int_{s}^t \hat{\eta_k}(t') dt' = \hat{\mathcal{N}}(0,t-s)$, where $\hat{\mathcal{N}}(0,1)$ is complex random variable that follows $\mathcal{N}(0,\frac{1}{2}) + i\mathcal{N}(0,\frac{1}{2})$ (See the supplemental text B). 
We then obtained the differential equation for $\hat{h_k}(t)$ for $k \geq 1 $ from \eqref{eq06} as follows:

\begin{equation} \label{eq07}
\frac{d}{dt} \hat{h_k} (t) = \hat{f_k} - \left( 4 \rho \sin^2{\frac{\pi k}{N}} \right) \hat{h_k}(t)+\sigma \hat{\eta_k}(t).
\end{equation}
$\hat{h}_{k}(t)$ was explicitly derived by using Ito integral as follows (See the supplemental text C):
\begin{equation} \label{eq08}
\hat{h}_{k}(t) = \frac{\hat{f_k}}{a_k} (1-e^{-a_k t}) - \sigma \int_{0}^{t} e^{a(s-t)} d \hat{B}_s. 
\end{equation}
$\hat{B}_{s}$ is Brownian motion in the complex plane, and $a_k = 4 \rho \sin^2{\frac{\pi k}{N}}.$
Equation \eqref{eq08} indicates that the Fourier coefficient $\hat{h}_k(t)$ follows the complex Gaussian distribution with the mean $\frac{\hat{f_k}}{a_k} (1-e^{a_k t})$ and the variance $\frac{\sigma^2}{2a_k} (1-e^{-2a_k t})$. The expected value of the power-spectrum $|\hat{h}_k(t)|^2$ is written as
\begin{equation} \label{eq09}
E[|\hat{h}_k(t)|^2] = \frac{|\hat{f}_k|^2}{a_k^2} (1-2 e^{-a_k t} +e^{-2 a_k t}) + \frac{\sigma^2}{2a_k} (1-e^{-2a_k t}). 
\end{equation}

\begin{figure}[hbt]
\includegraphics[width=8.6cm]{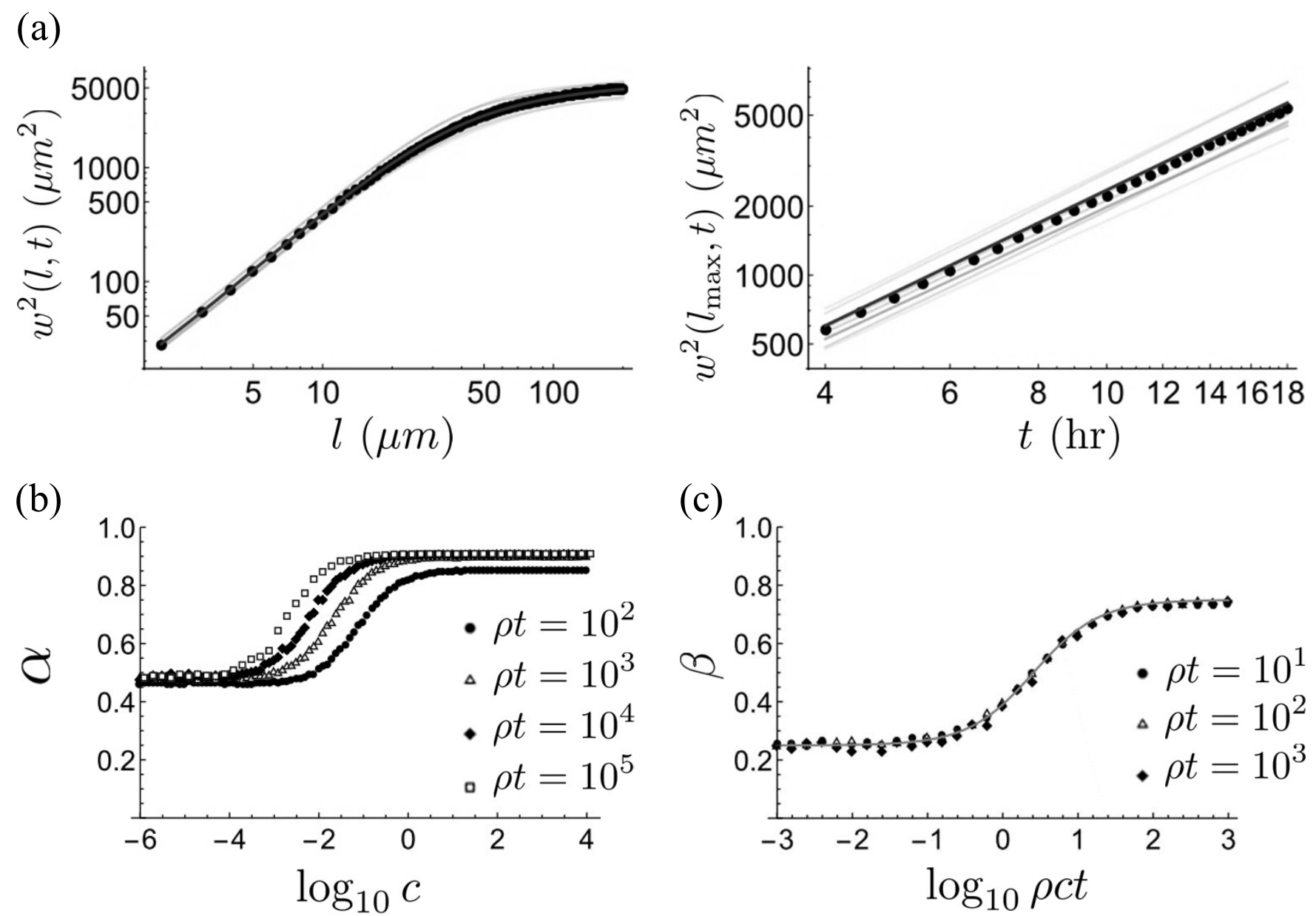}
\caption{\label{fig:5} Parameter dependency of the Hurst and growth exponents. 
(a) Comparison of numerically and analytically derived $w^2(l,t)$. Log-log plot of $w^2(l,t)$ against $l$ at $t = 18$ (left panel), and of $w^2(l_\text{max},t)$ against $t$ (right panel). The light-colored lines indicate the numerically obtained values with different sample paths, dots indicate the average of $w(l,t)$ for different paths, and the black line indicates analytically obtained values. (b) Numerically obtained parameter dependency of the Hurst exponent $\alpha$ on $c$ and $\rho t$. (c) The parameter dependency of the growth exponent $\beta$ on $\rho c t$ and $\rho t$. Dots indicate the numerically obtained growth exponents, and the line indicates the plot of \eqref{eq14}.}
\end{figure}

Next, we introduce the squared local roughness $w^2(l,t)$ as the average of the variance of $h_x$ in $m$ consecutive cells. Using the inter-cell distance $\Delta x$ and non-negative integer $m$, $w^2(l,t)$ is expressed as
\begin{equation} \label{eq10}
w^2(l,t)= w^2(m \Delta x,t) = R(0) - \frac{1}{m(m-1)}\sum_{l=1}^{m-1} 2(m-l) R(l).
\end{equation}
$R(l)$ is an auto-correlation function of $h_x$. Since $R(l)$ is obtained by inverse Fourier transform of the power spectrum $|\hat{h_k}(t)|^2$ (Winner-Khinchin's theorem), we obtained the expected value of $w^2(l,t)$ as in \eqref{eq10}.
Figure \ref{fig:5}(a) shows that the analytically derived $w^2(l,t)$ was close to the average of the numerical calculations of the circular model.

Considering the slope of the log-log plots in Fig. \ref{fig:5}(a), we obtained the Hurst and growth exponents as follows:
\begin{eqnarray}
\alpha &&= \log{\left[ 1+ \frac{1}{3} \frac{R(1)-R(2)}{R(0)-R(1)}\right]} / 2 \log{\frac{3}{2}} \label{eq11} \\
\beta &&= \frac{t}{2 w^2(l_\text{max},t)} \frac{\partial}{\partial t}w^2(l_\text{max},t). \label{eq12}
\end{eqnarray}

We calculated the expected values $\alpha = 0.80$ and $\beta = 0.74$ for the parameters used for the circular model in Fig. \ref{fig:4}. This confirmed that our analysis based on \eqref{eq06} captured the dynamics of the circular model in \eqref{eq03}. 

Equation \eqref{eq11} also shows that the Hurst exponent is represented as the ratio of the linear sum of the power spectra. Therefore, the multiplication of the power spectrum by a constant value should not affect the Hurst exponents. 
By dividing the power spectra \eqref{eq09} by $\sigma^2/\rho$, we found that the Hurst exponents were determined by the index $c$ and $\rho t$, where 
\begin{equation} \label{eq13}
c=\frac{(v_l-v_f)^2 p_l (1-p_l)}{\rho \sigma^2}.
\end{equation}
The plots of the Hurst exponents against $\log{c}$ with different $\rho t$ are shown in Fig. \ref{fig:5}(b). When the difference of cell motility ($v_l-v_f$) is relatively large, $c$ takes a large value and $\alpha$ approaches $\alpha = 0.91$, corresponding to the fixed noise model. On the other hand, when the random component of cell movement ($\sigma$) is relatively large, $c$ is small and $\alpha$ approaches $\alpha = 0.48$, corresponding to EW model.

The growth exponents $\beta$ were also regarded as a two-variable function of $\rho t$ and $c$. From the theoretical consideration (See the supplemental text D), we can calculate $\beta$ as follows:
\begin{equation} \label{eq14}
\beta = \frac{2 \left(2-\sqrt{2}\right) \gamma +1}{\frac{8}{3} \left(2-\sqrt{2}\right) \gamma +4}.
\end{equation}
where $\gamma = \rho c t$ and we assumed that $N$ and $\rho t$ are sufficiently large. Plots of $\beta$ against $\log{\gamma}$ are shown in Fig. \ref{fig:5}$(c)$. It was shown that the random cell movement decreases $\beta$. However, after sufficient time has passed, $\beta$ always takes a value of $0.75$, corresponding to fixed noise model.
\subsection{\label{sec:results5}Confirmation of analytical prediction}
As described in this section, we experimentally examined the dependency of the Hurst and growth exponents as derived from numerical and analytical considerations.

First, the effect of the initial size of the monolayer was examined. A monolayer of 3 mm diameter was prepared, and its spreading was observed. The Hurst and growth exponents were $\alpha = 0.837$ and $\beta = 0.740$, respectively, confirming that the scaling property of the growing contour was not dependent on the number of cells (Fig. S5). Therefore, this result was reproduced by the same parameters used in Fig. \ref{fig:2}. 

\begin{figure*}[hbt]
\includegraphics[width=17.8cm]{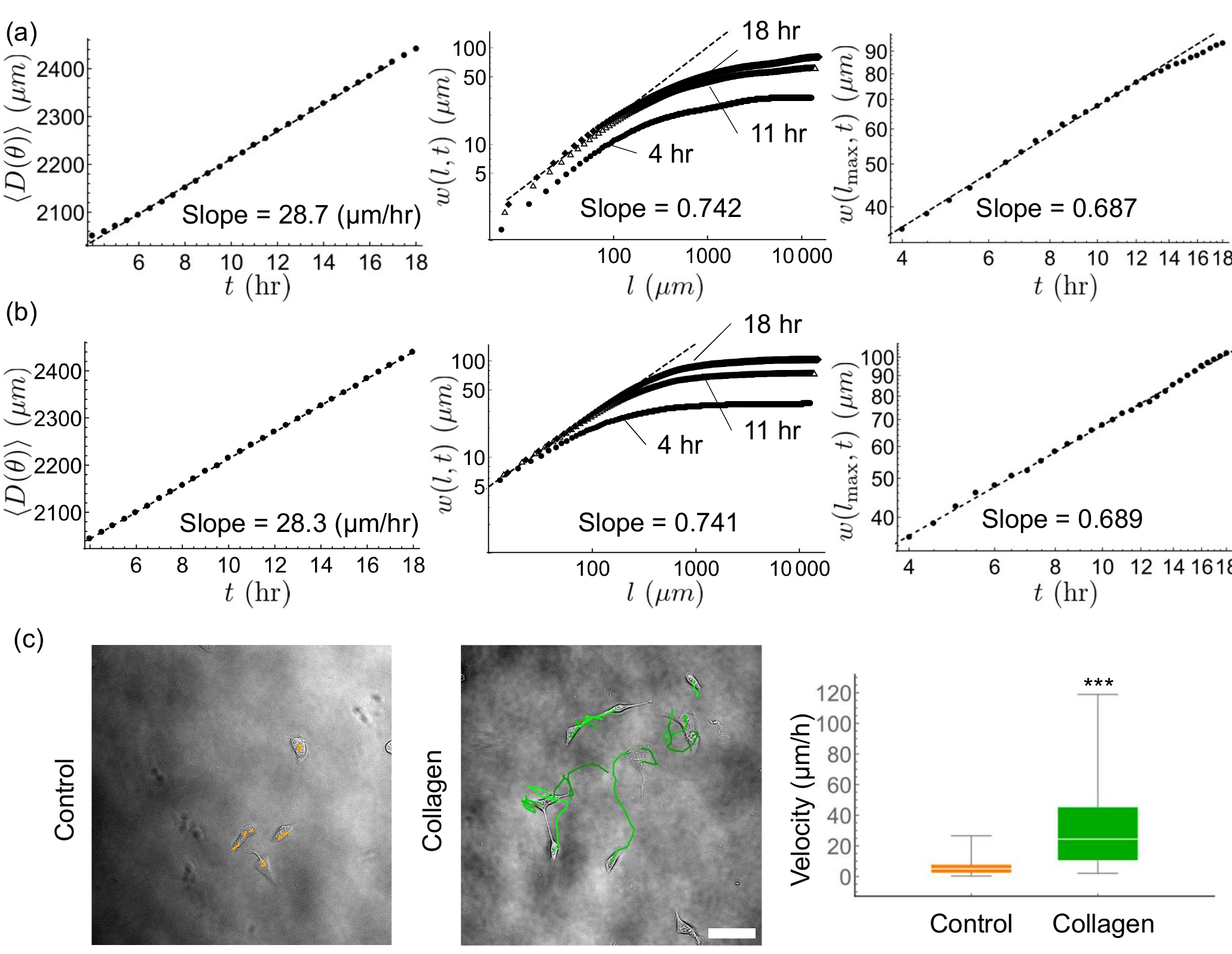}
\caption{\label{fig:6} The experimental result with $100 mg/L$ type I collagen coated dish is also reproduced by the model \eqref{eq02}. 
(a) Experimental results. Time evolution of averaged $D(\theta)$ (left panel), log-log plot of the local roughness $w(l,t)$ and $l$ (middle panel), and log-log plot of $w(l_\text{max},t)$ and $t$ (right panel). Dots indicate the experimentally obtained data and the dashed line is the fitted line for the data. (b) Numerical results corresponding to (a). Parameters: $v_l = 126 \ [\mu m\ h^{-1}]$, $v_f = 17.87 \ [\mu m \ h^{-1}]$, $p_l =0.1$, $\rho = 6.67 [h^{-1}]$, and $\sigma=28.8 \ [\mu m \ h^{-1/2}]$. (c) Single cell movement assay. The trajectories of the cell nuclei are indicated by the colored lines in the uncoated glass dish (left panel) and the collagen coated dish (middle panel). Right panel shows the velocity of the cells; $6.36 \mu m/h$ in the uncoated glass dish and $30.1 \mu m/h$ in the collagen coated dish. *** p-value = $3.68 \times 10^{-9}$, scale bar = $100 \mu m$.}
\end{figure*}

Next, we investigated how cell behavior affects the scaling property. Since it is known that cell behavior changes depending on the substrate \cite{Haga_Collective_2005, Prez-Gonzlez_Active_2019}, we prepared a collagen coated culture dish and performed time-lapse observations of the MDCK cell monolayer from $t=4$ to $t=18$.
The results of these observations are shown in Fig. \ref{fig:6}(a) and Supplemental Movie 4. We found that the Hurst and growth exponents both decreased: $\alpha = 0.742, \beta = 0.687$. Meanwhile, the averaged expansion speed of $\langle D(\theta) \rangle$ was increased to $28.7 \mu m$, which is $2.5$ times higher than on uncoated glass dish. In addition, we repeatedly measured the exponents in different monolayers, and obtained that $\alpha = 0.749 \pm 0.012$ and $\beta = 0.687 \pm 0.050$, respectively. These values were not much different from the data shown in Fig. \ref{fig:6}(a).

Using the results of the mathematical analysis (Fig. \ref{fig:5}(b), \ref{fig:5}(c)), we estimated the parameters that would reproduce the experimental data in Fig. \ref{fig:6}(a). We derived the value of $\gamma = \rho c t$ such that the growth exponent $\beta = 0.687$ was satisfied in \eqref{eq14}, and found $\gamma = 17.8$.
Here, we assumed that while the difference of the substrate affects the motility of the cell, the effects on cell-cell adhesion and leader cell emergence are small. Thus, the values of $\rho$ and $p_l$ are regarded as identical to those in the control (Fig. \ref{fig:4}): $\rho=6.67, \  p_l=0.1$. We obtained $c= 0.148$ and could then determine the values of $\sigma^2$ and $(v_l-v_f)^2 p_l (1-p_l)$ to match the values of $w(l_\text{max},t)$. 
In addition, the time evolution of the mean diameter is $23.8\ \mu m/h$, which is equal to $v_l p_l + v_f(1-p_l)$, so both $v_l$ and $v_f$ can be estimated. The values for the estimated parameters were $v_l=123, \  v_f=18.2$ and $\sigma=31.8$.

At the same time, the parameters can also be estimated by the Hurst exponents using the relationship in Figure \ref{fig:5}(b). We estimated that $c = 0.19$ from $\rho t = 120$ and $\alpha = 0.749$. The parameter values are obtained in the same manner: $v_l=126, \ v_f=17.9$ and $\sigma=28.8$.
The estimated parameter sets from Hurst and growth exponents take close values.
This result shows that the Hurst and growth exponents under the different conditions are also consistent with our model.
Figure \ref{fig:6}(b), S6 and Supplemental Movie 5 show the results of numerical calculations using the parameters $v_l=126, v_f=17.9$ and $\sigma=28.8$. These estimated parameters reproduced the experimental results.

Next, we confirmed the consistency between theoretically estimated and experimentally observed cell motility.
The experimental results (shown in Fig. \ref{fig:6}(a)) were reproduced in our model by assuming that the motility of the follower cell ($v_f$) and the intensity of random cell movement ($\sigma$) both increased about 9-fold from the parameters used in Fig. \ref{fig:2}.
Figure \ref{fig:6}(c) and Supplemental Movie 6 show the cell motilities on the different cell culture substrate. The total path lengths of the single cell for 2 hr were measured: $6.4 \ \mu m/h$ for the uncoated glass dish and $30.1 \ \mu m/h$ for the collagen coated dish. The motility of the cells on the collagen increase to $4.7$ folds. This value is not exactly matched to the theoretical prediction, however, the order of the values is not far off.
Therefore, the experimental results under the different conditions were consistent with the analytically predicted parameter dependency of the exponents, and the parameters to reproduce the experiment are quantitatively reasonable. These results further confirm that the contour formation of MDCK cells is generally well-explained by our model.

\section{\label{sec:discussion}Discussion}
In this study, we propose a concept that explains the mechanism of the dynamic scaling law observed in the spreading of the epithelial cell monolayer.
We found that the time evolution of the contour satisfied the dynamic scaling law: $\alpha=0.86$ and $\beta = 0.73$. Based on the observation results, we constructed a simple mathematical model, and demonstrated that the contour shape arose from the behavior of the cells at the sheet edge. 
Our mathematical analysis of the model suggested that the presence of leader cells is essential for observed pattern formation. We found that the Hurst and growth exponents were dependent only on the ratio of the variance of cell motilities to the intensity of the random cell movement. The theoretical prediction was experimentally confirmed by changing the cell motilities. Thus, our study offers a new framework for examining dynamic scaling in the biological phenomenon.

The EW and KPZ models are widely known to satisfy the dynamic scaling law, and it has been reported that the contour shape of HeLa and Vero cell colonies follows KPZ universality. However, the MDCK cells observed here did not follow this universality. Possible reasons for this difference are the emergence of leader cells and the differences in observation time.
First, the leader cells emerge in the early stage and persist during observation, and then the differential motility corresponds to the fixed noise in the model. On the other hand, the EW and KPZ equations do not contain the corresponding noise term. In addition, the effect of the nonlinear component in the KPZ system requires a long observation time. In the report by Huergo and colleagues, the observation time was 13000 minutes \cite{Huergo_Growth_2012}, while the observation time in this study was 1080 minutes.

We focused on the contour shape change that emerged from cell motility and cell-cell adhesion. In the view of biology, the model used in this study suggests that cell motility played an important role in the contour formation, while the effect of cell proliferation would be dominant after a long period of time. From a model perspective, the increase in the number of cells on the circumference is not taken into account. Therefore, the model cannot explain the pattern formation in a long period of time that the effect of proliferation on the contour shape cannot be negligible.

The growth exponents obtained from analysis of the mathematical model, $\beta = 0.74$, were almost identical to those obtained numerically and experimentally. While the Hurst exponent, $\alpha=0.80$,was smaller than those obtained by numerical calculations $\alpha = 0.85$. The value of $w(l,t)$ is defined as the mean of the standard deviation in the closed range $l$, however, this value cannot be directly calculated from the power spectra. Therefore, to estimate the Hurst exponent, the expected value of variance was calculated, and the square root of this value was taken to estimate the value of $w(l,t)$. 
The value of $\sqrt{w^2(l,t)}$ is not the same as $w(l,t)$, since the mean of the standard deviation in a given interval is different from the square root of the mean of the variance. On the other hand, for the growth exponents, we calculate global roughness $w(l_\text{max},t)$ as the standard deviation with respect to the whole direction. Since this value is equal to the square root of the variance, it is close to the values obtained numerically.

The study of mathematical models of collective cell movement has attracted significant attention over the past decade, especially within the fields of statistical mechanics and biophysics. The models can be categorized into continuous models and discrete models. In continuous models, the cell colony is regarded as a continuum. Such models have been constructed mainly to explain the fingering instability of the epithelial sheet \cite{Ouaknin_Stochastic_2009, Mark_Physical_2010, Lee_Crawling_2011, Kpf_A_2013, Prez-Gonzlez_Active_2019, Alert_Active_2019}.


On the other hand, in discrete models, each cell has been represented by polygons or particles. In the former case, the classical vertex model with chemotaxis and fluid properties \cite{Salm_Chemical_2012} and the active vertex model \cite{Barton_Active_2017}, which also added the effect as an active fluid, have been proposed. In the latter, a particle model mostly included the cell-cell interactions and the random kinetic components. The model explicitly introduced leader cells \cite{Seplveda_Collective_2013}, and a model with the effects of the bending and the surface tension have been proposed \cite{Tarle_Modeling_2015}.
These discrete models tend to be descriptive, and assume many factors that affect cell behavior. While such models are preferable to explain the experimental data, however, it is likely to be difficult to analytically explain the numerical results due to the model complexity. Thus, analytical considerations of the model equation are required to fully understand the relationship between the physical quantity and the scaling property. 

In a sense, our model can be understood as the simplest form of the discrete particle model. The reason we used the spatially discrete model was to describe the differences of the motilities among the cells, such as leader or follower cells. The result that even the simple model explained the dynamic scaling within the contour shape suggested that the random cell movement, the deterministic differences in cell motility, and the effects of intercellular adhesion have critical implications on the contour of the epithelial monolayer.

The expected values of the power spectra \eqref{eq09} converge to the constant values when $\rho t \to \infty$. Since $w^2(l,t)$ are represented by the linear sum of the power spectra, $w^2(l,t)$ also converges to constant values. Therefore, it is suggested that $\alpha$ at $\rho t \to \infty$ is determined by $c$, and that there is a critical value $c^*$ such that when $c \ll c^*$, the scaling law $\alpha=0.5, \beta=0.25$ was obtained and when $c \gg c^*$, $\alpha=0.9, \beta=0.75$ was obtained. The value of $c^*$ was estimated from the analytically obtained relationship between expected values of $w^2(l,t_\infty)$.

The equation known as quenched Edwards-Wilkinson (QEW) equation \cite{Kessler_interface_1991,Meakin_Fractals_2011} is a model with the noise term dependent on $x$ and $h$ and the driving force $\mathcal{F}$ as follows:
\begin{equation} \label{eq15}
  \frac{\partial}{\partial t} h(x, t)=\rho \nabla^{2} h(x, t)+\mathcal{F}+\eta(x, h(x, t)).
\end{equation}
In this model, when $\mathcal{F}$ is sufficiently large, the noise term is considered to be spatially and temporally independent, then, the system could be identified as the EW equation: $\alpha=0.5, \beta=0.25$. However, the transition to a different dynamic scaling law $\alpha=1.0$ and $\beta=0.75$ occur when $\mathcal{F} \ll \mathcal{F}_c$ \cite{Kessler_interface_1991,Narayan_Threshold_1993,Meakin_Fractals_2011}. 
The relationship between $\mathcal{F}_c$ and the scaling law is similar to the relationship between $c^*$ and the scaling law in our model \eqref{eq06}, which suggests that our models could possibly be identified as QEW-type models. However, we expect that more complex theoretical methods will be needed to solve this problem, and it remains a topic for future research.

\begin{acknowledgments}
We acknowledge fruitful conversations with S. Ishihara (University of Tokyo), T. Ogawa (Meiji University) and H. Yasaki (Meiji University). 
This work has been supported by JSPS through Grants No. JP18K06260 to H.T.I.

\end{acknowledgments}

\bibliography{readcube_export}
\bibliographystyle{apsrev4-1}
\end{document}


\section{Mathematical procedures}
\subsection{\label{app:a} Distribution of $\hat{f_x}$}
The cell motility $f_x$ takes the higher value $v_l$ with the probability $p_l$ or the lower value $v_f$ with $1-p_l$.
The discrete Fourier transformation of $f_x$ is represented as follows:
\begin{equation}
\begin{aligned}
\hat{f}_{k} &=\frac{1}{\sqrt{N}} \sum_{x=1}^{N} f_x e^{\frac{i 2\pi k x}{N}} \\
&=\frac{1}{\sqrt{N}} \sum_{x=1}^{N} (v_f+ (v_l-v_f)X_{x}) e^{\frac{i 2 \pi k x}{N}} \\
&=(v_l-v_f) \frac{1}{\sqrt{N}} \sum_{x=1}^{N} X_{x} e^{\frac{i 2 \pi k x}{N}} \\
&=(v_l-v_f) \frac{1}{\sqrt{N}} \sum_{x=1}^{N} \left(\cos \frac{2 \pi k x}{N}+i \sin \frac{2 \pi k x}{N}\right) X_{x},
\end{aligned}
\end{equation}
where $X_x$ is a random variate that takes $0$ or $1$ with the probabilities $p_l$ or $1-p_l$, respectively. The real part of $\hat{f}_k$ is represented as follows:
\begin{equation}
\text{Re}[\hat{f_k}]= (v_l-v_f) \frac{1}{\sqrt{N}} \sum_{x=1}^{N} X_{x} \cos{\frac{2 \pi k x}{N}}.
\end{equation}
The expected value $E[\text{Re}[\hat{f_k}]] = 0$ from $P(X_x=1) = P(X_{N-x}=1)$.
The variance of $\text{Re}[\hat{f_k}]$ is represented as follows:
\begin{equation}
\begin{aligned}
\text{Var}[\text{Re}[\hat{f_k}]] &= (v_l-v_f)^2 \frac{1}{N} E \left[ \left( \sum_{x=1}^N X_x \cos{\frac{2\pi kx}{N}} \right)^2 \right] \\
&=(v_l-v_f)^2 \frac{1}{N} E\left[\sum_{y=1}^{N} \sum_{x=1}^{N} X_{x} \cdot X_{y} \cos \frac{2 \pi k x}{N} \cos \frac{2 \pi k y}{N}\right]\\
&=(v_l-v_f)^2 \frac{1}{N}\left(\sum_{y=1}^{N} \sum_{x=1}^{N} E\left[X_{x} X_{y}\right] \cos \frac{2 \pi k x}{N} \cos \frac{2 \pi k y}{N}\right)\\
&=(v_l-v_f)^2 \frac{1}{N}\left(p_l \sum_{x=1}^{N} \frac{1}{2}\left(1-\cos \frac{4 \pi k x}{N}\right)  
-p_l^{2} \sum_{x=1}^{N} \frac{1}{2}\left(1-\cos \frac{4 \pi kx}{N}\right)\right)\\
&=\frac{1}{2} (v_l-v_f)^2 p_l(1-p_l),
\end{aligned}
\end{equation}
where
\begin{equation}
E[X_x X_y] = 
\left[ 
\begin{array}{ll}
p & \text{(if $x=y$)}\\
p^2 & \text{(otherwise)},
\end{array}
\right.
\end{equation}
and 
\begin{eqnarray}
\sum_{x=1}^N \sum_{y=1}^N \cos{\frac{2\pi kx}{N}} \cos{\frac{2\pi ky}{N}} =  
\left[ 
\begin{array}{cc}
\frac{1}{2}\left(1-\cos \frac{4 \pi kx}{N}\right) & \text{(if $x=y$)}\\
-\frac{1}{2}\left(1-\cos \frac{4 \pi kx}{N}\right) & \text{(if $x=N-y$)}\\
0 & \text{(otherwise)}, 
\end{array}
\right.
\end{eqnarray}
from the orthogonality of trigonometric function.
The imaginary part of $\hat{f}_k$ is also represented as follows:
\begin{eqnarray}
E[\text{Im}[\hat{f_k}]] &=& 0, \\
\text{Var}[\text{Im}[\hat{f_k}]] &=& \frac{1}{2} (v_l-v_f)^2 p_l(1-p_l).
\end{eqnarray}
Therefore, $\hat{f}_k$ follows the complex Gaussian distribution denoted as $(v_l-v_f) \sqrt{p_l(1-p_l)} \hat{\mathcal{N}}(0,1)$.
\subsection{\label{app:b} Distribution of $\hat{\eta}_k(t)$}
The discrete Gaussian white noise $\eta_{x} (t)$ is defined as follows: 
\begin{equation}
    B_x(t) = \int_0^{t} \eta_x(s) ds \sim \mathcal{N}(0,t)
\end{equation}
where $B_x(\cdot)$ is the Brownian motion. 
The discrete Fourier transformation of $B_x(t)$ is represented as follows:
\begin{equation}
\begin{aligned}
\hat{B}_k (t) &=\frac{1}{\sqrt{N}} \sum_{x=1}^{N} B_x(t) e^{\frac{i 2 \pi k x}{N}}\\
&=\frac{1}{\sqrt{N}} \sum_{x=1}^{N}\left(\cos \frac{2 \pi k x}{N}+i \sin \frac{2 \pi kx}{N}\right) B_x(t).
\end{aligned}
\end{equation}
The real part of $\hat{B}_k (t)$ is denoted by $\mathcal{R}_k(t)$ as follows,
\begin{equation}
\mathcal{R}_{k}(t)=\frac{1}{\sqrt{N}} \sum_{x=1}^{N} \cos{\frac{2\pi kx}{N}} B_x(t).
\end{equation}
The stochastic process $\mathcal{R}_k(t)$ is a Gaussian process with the mean = $0$, since $\mathcal{R}_{k}(t)$ is the weighted average of $B_x(t)$.

Next, the variance of $\mathcal{R}_k(t)$ is represented as follows:
\begin{equation}
\begin{aligned}
\text{Var}\left[\mathcal{R}_{k}(t)\right] &=\frac{1}{N} E\left[\mathcal{R}_{k}(t)^{2}\right] \\
&=\frac{1}{N} E\left[ \sum_{x=1}^{N} \sum_{y=1}^{N} \cos{\frac{2 \pi k x}{N}} B_{x}(t) \cos \frac{2 \pi k y}{N} B_y(t)\right] \\
&=\frac{1}{N} \sum_{x=1}^{N} \cos^{2}{\frac{2 \pi k x}{N}} E\left[ B_x^{2}(t) \right] \\
&= \frac{1}{2} t.
\end{aligned}
\end{equation}
Therefore, the probability distribution of $\mathcal{R}_k(t)$ follows the normal distribution, with mean =  $0$ and variance = $t/2$.

Similarly, the imaginary part of $\hat{B}_k (t)$ is denoted $\mathcal{I}_k(t)$,
\begin{equation}
\mathcal{I}_{k}(t)=\frac{1}{\sqrt{N}} \sum_{x=1}^{N} \sin{\frac{2\pi kx}{N}} B_x(t).
\end{equation}
$\mathcal{I}_k(t)$ is also a Gaussian process with a mean of $0$, and
\begin{equation}
\begin{aligned}
\text{Var}\left[\mathcal{I}_{k}(t)\right] &=\frac{1}{N} \sum_{x=1}^{N} \frac{1}{2}\left(1-\cos \frac{4 \pi k x}{N}\right) t \\
&= \frac{1}{2} t.
\end{aligned}
\end{equation}
$\hat{\eta}_k(t)$ is represented as the derivative of $\hat{B}_k(t)$ at $t$. Therefore, the discrete Fourier transformation of $\eta_x(t)$ is represented as follows:
\begin{equation}
\hat{\eta}_k(t) = \frac{1}{\sqrt{2}}(\eta_k(t) + i \eta_k(t)).
\end{equation}
\subsection{\label{app:c} Expected value of power spectra $|\hat{h}_k|^2$ }
The stochastic process $X_{k,t} = \hat{f_k}- a_k \hat{h}_{k,t}$ satisfies the following equation from (7):
\begin{equation}
d X_{k,t}=-a_k X_{k,t} dt - a_k \sigma d \hat{B}_{t},
\end{equation}
where $a_k = 4 \rho \sin^2{\frac{\pi k}{N}}$. 
The stochastic process $Z_{k,t} = X_{k,t} e^{a_k t}$ satisfies the following equation:
\begin{equation}
\frac{d Z_{k,t}}{d t}= a_k X_{k,t}  e^{a_k t}, \quad \frac{d Z_{k,t}}{d X_{k,t}}= e^{a_k t}, \quad \frac{d^2 Z_{k,t}}{d X_{k,t}^2}= 0 .
\end{equation}

By using Ito's lemma, $Z_{k,t}$ is represented as follows:
\begin{equation}
d Z_{k,t} = a_k X_{k,t} e^{a_k t} dt + e^{a_k t} dX_{k,t} =a_k \sigma e^{a_k t} d \hat{B}_{t}.
\end{equation}
Therefore, $X_{k,t}$ is represented as follows:
\begin{equation}
X_{k,t} = X_{k,0} e^{-a_k t} +a_k \sigma \int_{0}^{t} e^{a_k (s-t)} d \hat{B}_s .
\end{equation}
The values of $X_{k,t} = \hat{f_k}- a_k \hat{h}_{k,t}$ and $h_{k,x}$ are constant when $t=0$, thus we obtain $X_{k,0} = \hat{f_k}$. Therefore, $\hat{h}_{k,t}$ is represented as follows:
\begin{equation}
\hat{h}_{k,t} = \frac{\hat{f_k}}{a_k} (1-e^{-a_k t}) - \sigma \int_{0}^{t} e^{a_k (s-t)} d \hat{B}_s,
\end{equation}
where the first term in right-hand side is a deterministic process, and the second term is the Gaussian process with mean $0$. From ${|d \hat{B}_s|}^2 \sim ds$, the variance of $\hat{h}_{k,t}$ is represented as follows:
\begin{equation}
\text{Var}[\hat{h}_{k,t}] = \sigma^2 e^{-2a_k t} \int_{0}^t e^{2a_k s} ds = \frac{\sigma^2}{2 a_k} (1-e^{-2a_k t}).
\end{equation}
The Fourier coefficient $\hat{h}(t)$ follows the complex Gaussian distribution with the mean $\frac{\hat{f_k}}{a_k} (1-e^{a_k t})$ and the variance $\frac{\sigma^2}{2a_k} (1-e^{-2a_k t})$.

Therefore, the expected value of the power spectra  $|\hat{h}_k(t)|^2$ is represented as follows:
\begin{equation}
E[|\hat{h}_k(t)|^2] = \frac{|\hat{f}_k|^2}{a_k^2} (1-2 e^{-a_k t} +e^{-2 a_k t}) + \frac{\sigma^2}{2a_k} (1-e^{-2a_k t}).
\end{equation}

\subsection{\label{app:d} Deriving the growth exponents}
From equation (10), the growth exponent $\beta$ is represented as follows:
\begin{equation} \displaystyle \label{beta}
\beta = \frac{c \ f(\tau)  +\  g(\tau)}{c \tau \ h(\tau)+ \tau \ i(\tau)},
\end{equation}
where
\begin{eqnarray} \displaystyle
f(\tau) &=&  \frac{1}{N} \sum_{k=1}^N \frac{1-2 e^{-s^2 \tau} +e^{-2s^2 \tau}}{s^4},  \\
g(\tau) &=&  \frac{1}{N} \sum_{k=1}^N \frac{1-e^{-2s^2 \tau}}{s^2},  \\
h(\tau) &=&  \frac{1}{N} \sum_{k=1}^N \frac{2 e^{-s^2 \tau}-2 e^{-2s^2 \tau}}{s^2},  \\
i(\tau) &=&  \frac{1}{N} \sum_{k=1}^N e^{-2s^2 \tau}, 
\end{eqnarray} 
and we set $\tau = \rho t$ and $s = 2\sin{(\pi k/N)}$.

Assuming that $N$ is sufficiently large, $f(x), g(x), h(x)$ and $i(x)$ are approximated by the integral as follows:
\begin{eqnarray} \displaystyle
f(\tau) &\simeq&  \int_{0}^{1} \frac{2 \left(e^{-2\tau  (2\pi x)^2}-2 e^{-\tau (2\pi x)^2}+1\right)}{(2\pi x)^4} dx  \\
g(\tau) &\simeq&  \int_{0}^{1} \frac{2 \left(1-e^{-2 \tau  (2 \pi x)^2}\right)}{(2\pi x)^2} dx  \\
h(\tau) &\simeq&  \int_{0}^{1} \frac{4 \left(e^{-\tau  (2 \pi x)^2}-e^{-2 \tau  (2 \pi x)^2}\right)}{(2 \pi x)^2} dx  \\
i(\tau) &\simeq&  \int_{0}^{1} e^{- (2\pi x)^2  \tau} dx.
\end{eqnarray}

Assuming that $\tau$ is sufficiently large, each of the integrals is simplified as follows:
\begin{eqnarray} \label{integral} \displaystyle
f(\tau) &\simeq&  -\frac{4 \left(1-\sqrt{2}\right) \tau ^{3/2}}{3 \sqrt{\pi }}  \\
g(\tau) &\simeq&  \sqrt{\frac{2}{\pi }} \sqrt{\tau }  \\
h(\tau) &\simeq&  \frac{2 \left(\sqrt{2}-1\right) \sqrt{\tau }}{\sqrt{\pi }}  \\
i(\tau) &\simeq&  \frac{1}{\sqrt{2 \pi } \sqrt{\tau }}.
\end{eqnarray}

Therefore, the growth exponents $\beta$ are determined by the index $\gamma = c \tau$. Substituting these into (\ref{beta}), $\beta$ are represented as follows:
\begin{equation}
\beta = \frac{2 \left(2-\sqrt{2}\right) \gamma +1}{\frac{8}{3} \left(2-\sqrt{2}\right) \gamma +4}
\end{equation}

\newpage

\section{Supplemental movies and figures}

\subsection{Movie 1: Spreading of the MDCK cell monolayer on the uncoated glass dish corresponding to Fig. 2(a)}
The MDCK cells were visualized by CellTracker Green, which localized to the cytoplasm.
The region with the green fluorescent indicates the MDCK cell monolayer.
Images were taken with $30$ min intervals, from $t=4$ hr to $18$ hr.
Scalebar = $1000 \ \mu m$

\subsection{Movie 2: Tracking of the cell nuclei at the edge of monolayer}
Magenta circles indicate the cell nuclei that are at edge of the monolayer at $t=1$ hr and $18$ hr.
Cells were tracked manually with TrackMate plugin.
Paths were colored by the order of the mean speed (red lines indicate the paths with the high mean speed, and blue lines indicate the paths with the low mean speed).
Images were taken with $20$ min intervals, from $t=1$ hr to $18$ hr.
Scalebar = $200 \ \mu m$

\subsection{Movie 3: Numerical simulation of the spreading of MDCK monolayer, corresponding to Fig. 4}
The dynamics of the cells at the edge of the monolayer were simulated, and the shape of the monolayer was estimated by linking the cells.
Parameters: $v_l = 94 \ [\mu m\ h^{-1}]$, $v_f = 2 \ [\mu m \ h^{-1}]$, $p_l =0.1$, $\rho = 6.67 [h^{-1}]$, and $\sigma=3 \ [\mu m \ h^{-1/2}]$.

\subsection{Movie 4: Spreading of the MDCK cell monolayer on the collagen-coated dish}
The MDCK cells were visualized by CellTracker Green and the region with the green fluorescent indicates the MDCK cell monolayer.
Images were taken at $30$ min intervals, from $t=4$ hr to $18$ hr.
Scalebar = $1000 \ \mu m$

\subsection{Movie 5: Numerical simulation of the spreading of MDCK monolayer on the collagen-coated dish, corresponding to Fig. 6}
The dynamics of the cells at the edge of the monolayer were simulated, and the shape of the monolayer was estimated by linking the cells.
Parameters: $v_l = 126 \ [\mu m\ h^{-1}]$, $v_f = 17.87 \ [\mu m \ h^{-1}]$, $p_l =0.1$, $\rho = 6.67 [h^{-1}]$, and $\sigma=28.8 \ [\mu m \ h^{-1/2}]$

\subsection{Movie 6: Time-lapse observation of single cell motilities under the different substrate conditions, corresponding to Fig. 6}
Images were the bright field images and taken with $5$ min intervals for $2$ hr.
Scalebar = $100 \ \mu m$

\newpage

\begin{figure*}[hbt]
\includegraphics[width=10cm]{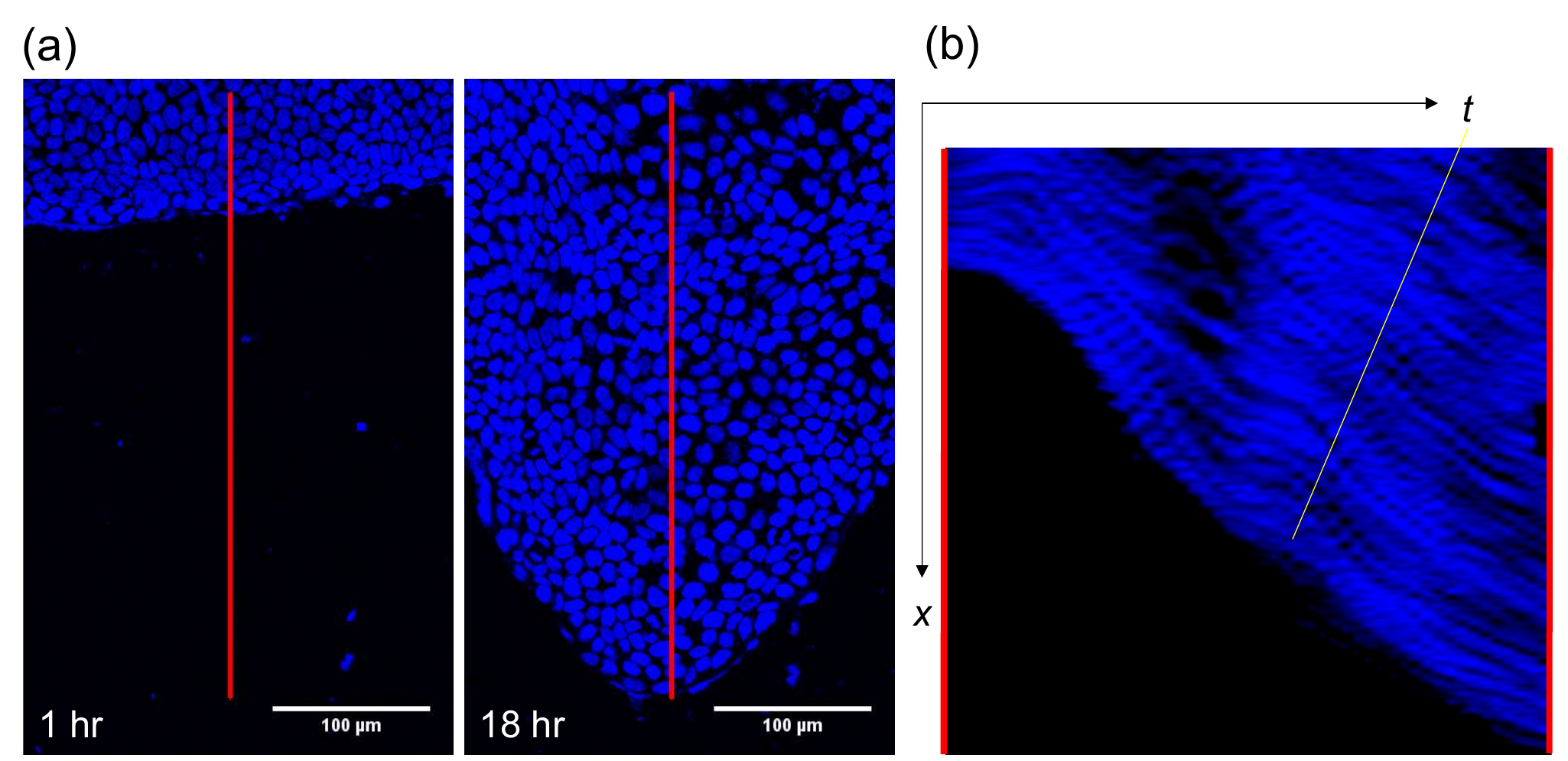}
\caption{\label{fig:12} Cell migration started from the cells at the edge.
(a) Cell nuclei were visualized by Hoechst 33342. The cells migrated to the outer regions. Red lines are the reference lines for making the kymograph. (b) Kymograph of cell migration. Images were constructed by stacking the fluorescent intensity along with the red lines from $t=1$hr (the right edge) to $t=18$ hr (the left edge). The yellow line indicates that the cell movement at the edge was propagated to the inner region.}
\end{figure*}

\begin{figure*}[hbt]
\includegraphics[width=10cm]{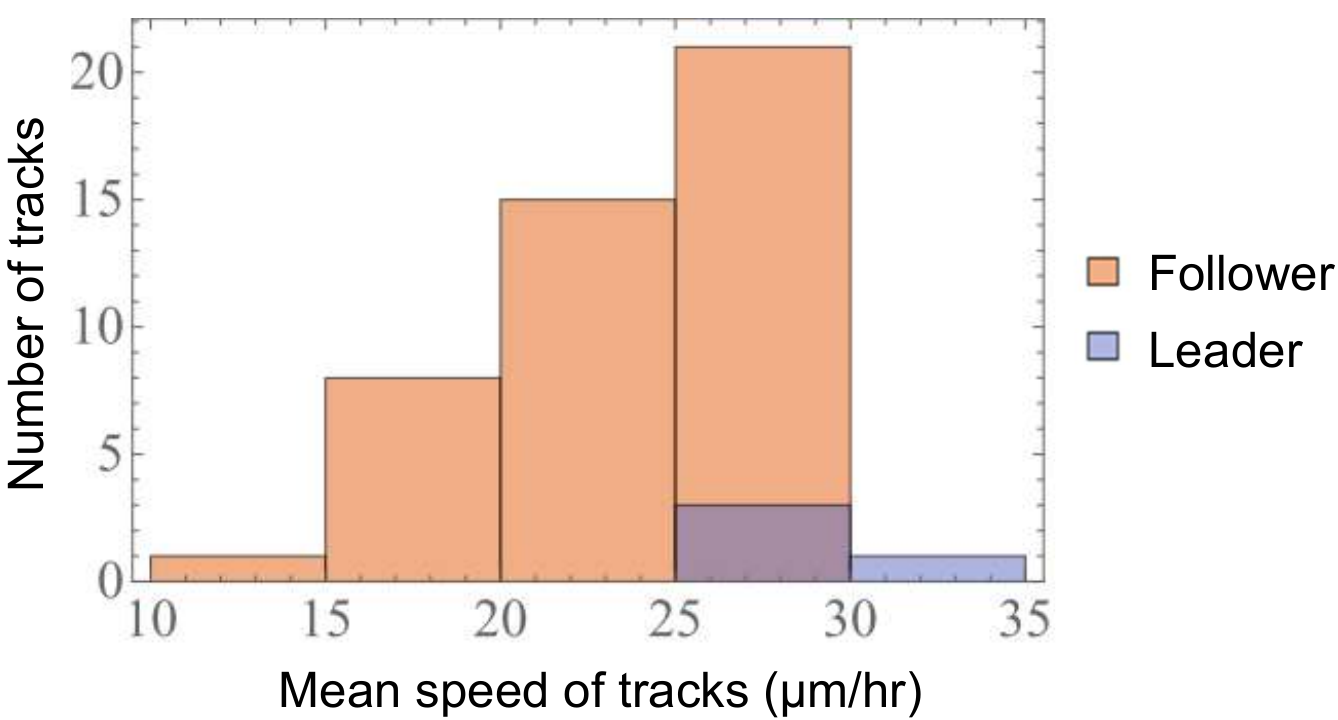}
\caption{\label{fig:11} Distribution of mean speed from cell tracking data corresponding to Fig. 3(a) and Movie 2. A red box indicates the tracks of the follower cells ($n=45$), and a blue box indicates the tracks of the leader cells ($n=4$).}
\end{figure*}

\begin{figure*}[hbt]
\includegraphics[width=15cm]{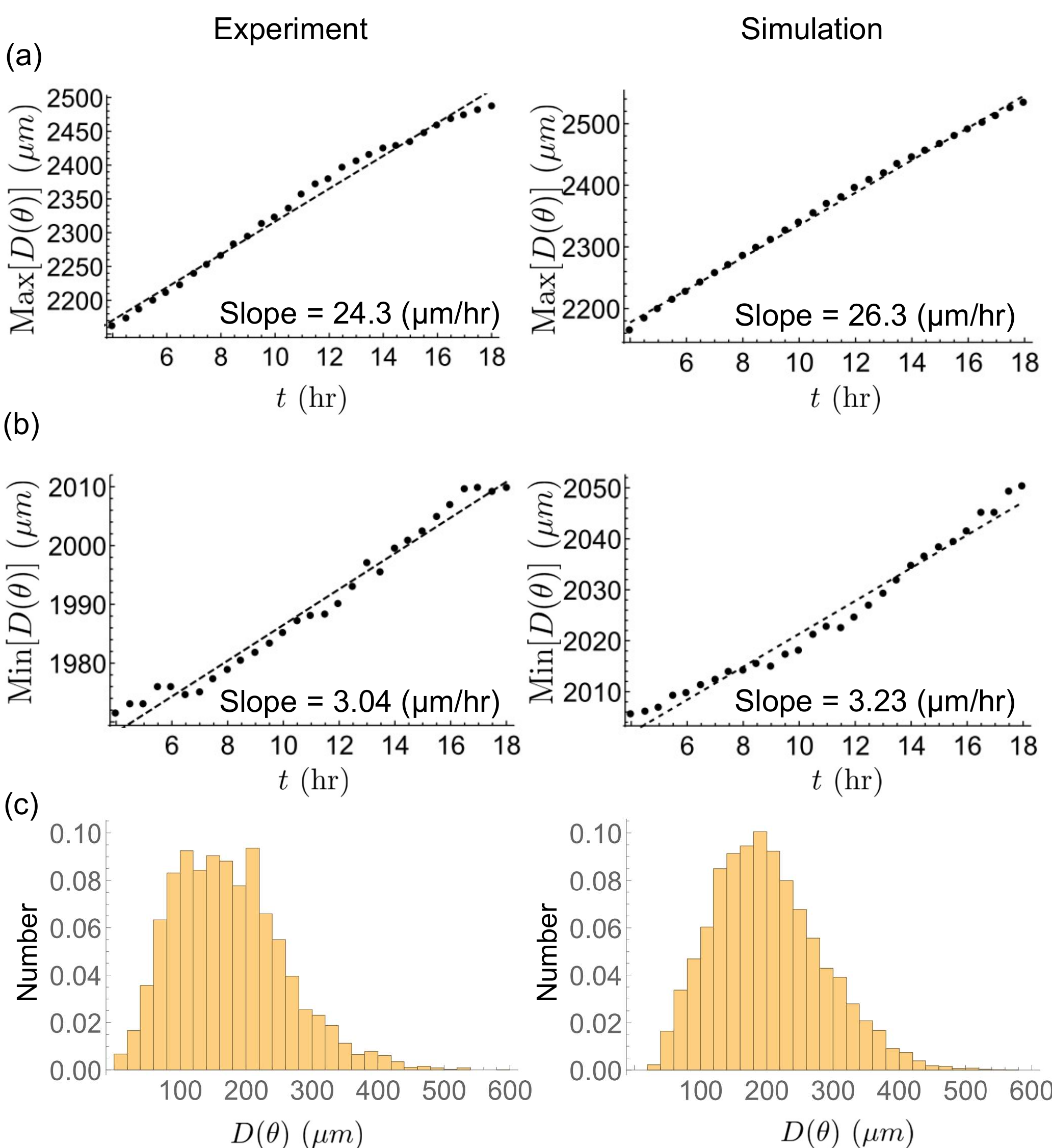}
\caption{\label{fig:7} Comparison of the experimentally and the numerically obtained results, corresponding to Fig. 2 and Fig. 4.
(a) Time evolution of the maximum values of $D(\theta)$. (b) Time evolution of the minimum values of $D(\theta)$. (c) Distribution of the increment of $D(\theta)$ in 18 hr.
Left panels show the experimental results ($n=6$), and right panels show the numerical results.
Parameters for numerical simulation: $v_l = 94 \ [\mu m\ h^{-1}]$, $v_f = 2 \ [\mu m \ h^{-1}]$, $p_l =0.1$, $\rho = 6.67 [h^{-1}]$, and $\sigma=3 \  [\mu m \ h^{-1/2}]$.}
\end{figure*}

\begin{figure*}[hbt]
\includegraphics[width=15cm]{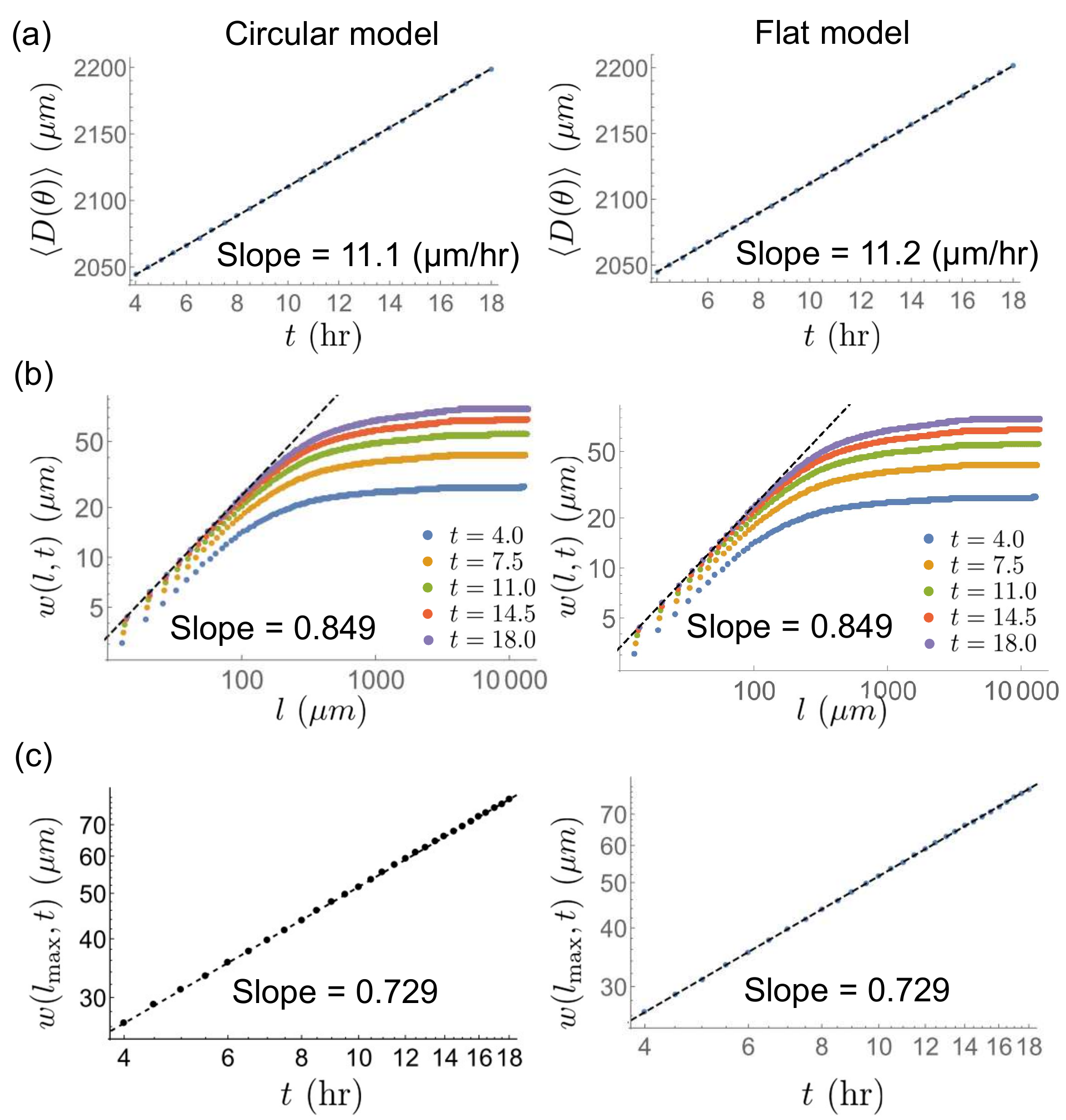}
\caption{\label{fig:8}Comparison of the circular model and the flat model.
(a) Time evolution of the averaged values of $D(\theta)$.
(b) Local roughness $w(l,t)$ was scaled for $l$ when $l$ is small, and the Hurst exponent was calculated: $\alpha = 0.849$ for each results.
(c) Global roughness $w(l_\text{max},t)$ was scaled for $t$, and the growth exponent was calculated: $\beta = 0.729$ for each results.
Left panels show the results obtained from the circular model, and the right panels show the results obtained from the flat model.
The parameters and the leader cell distribution used in each simulation are the same as Fig. 4.
Parameters: $v_l = 94 \ [\mu m\ h^{-1}]$, $v_f = 2 \ [\mu m \ h^{-1}]$, $p_l =0.1$, $\rho = 6.67 [h^{-1}]$, and $\sigma=3 \  [\mu m \ h^{-1/2}]$.}
\end{figure*}

\begin{figure*}[hbt]
\includegraphics[width=17.8cm]{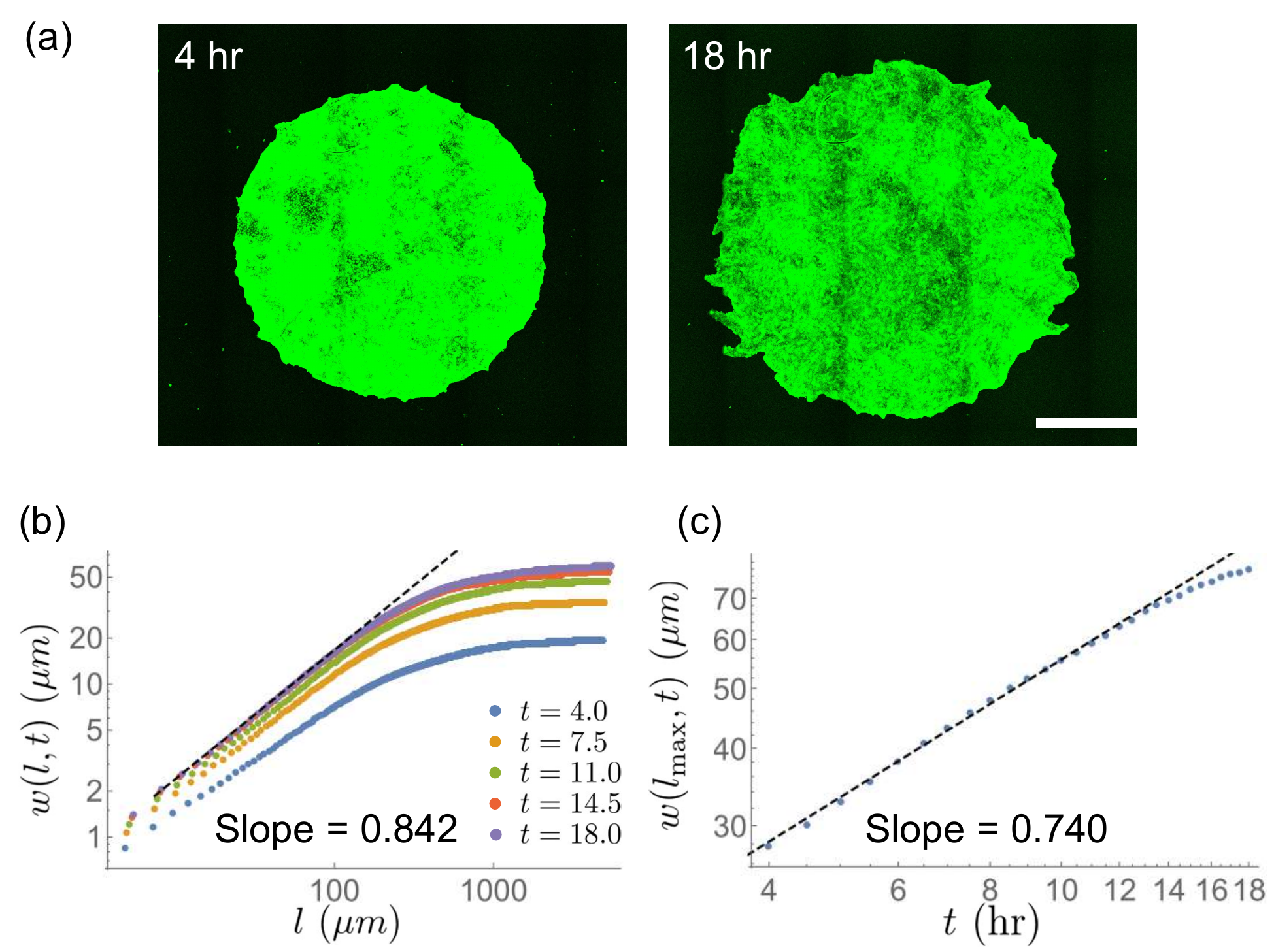}
\caption{\label{fig:9}Experimental results of the small size monolayer.
(a) The monolayer spread from the $3 mm$ diameter disk. The left panel shows the shape of the monolayer at $t = 4 hr$, and the right panel at $t= 18 hr$. Scalebar = 1000 $\mu m$.
(b) Local roughness $w(l,t)$ was scaled for $l$ when $l$ is small, and the Hurst exponent was calculated: $\alpha = 0.842$.
(c) Global roughness $w(l_\text{max},t)$ was scaled for $t$, and the growth exponent was calculated: $\beta = 0.740$ ($n=6$).}
\end{figure*}

\begin{figure*}[hbt]
\includegraphics[width=15cm]{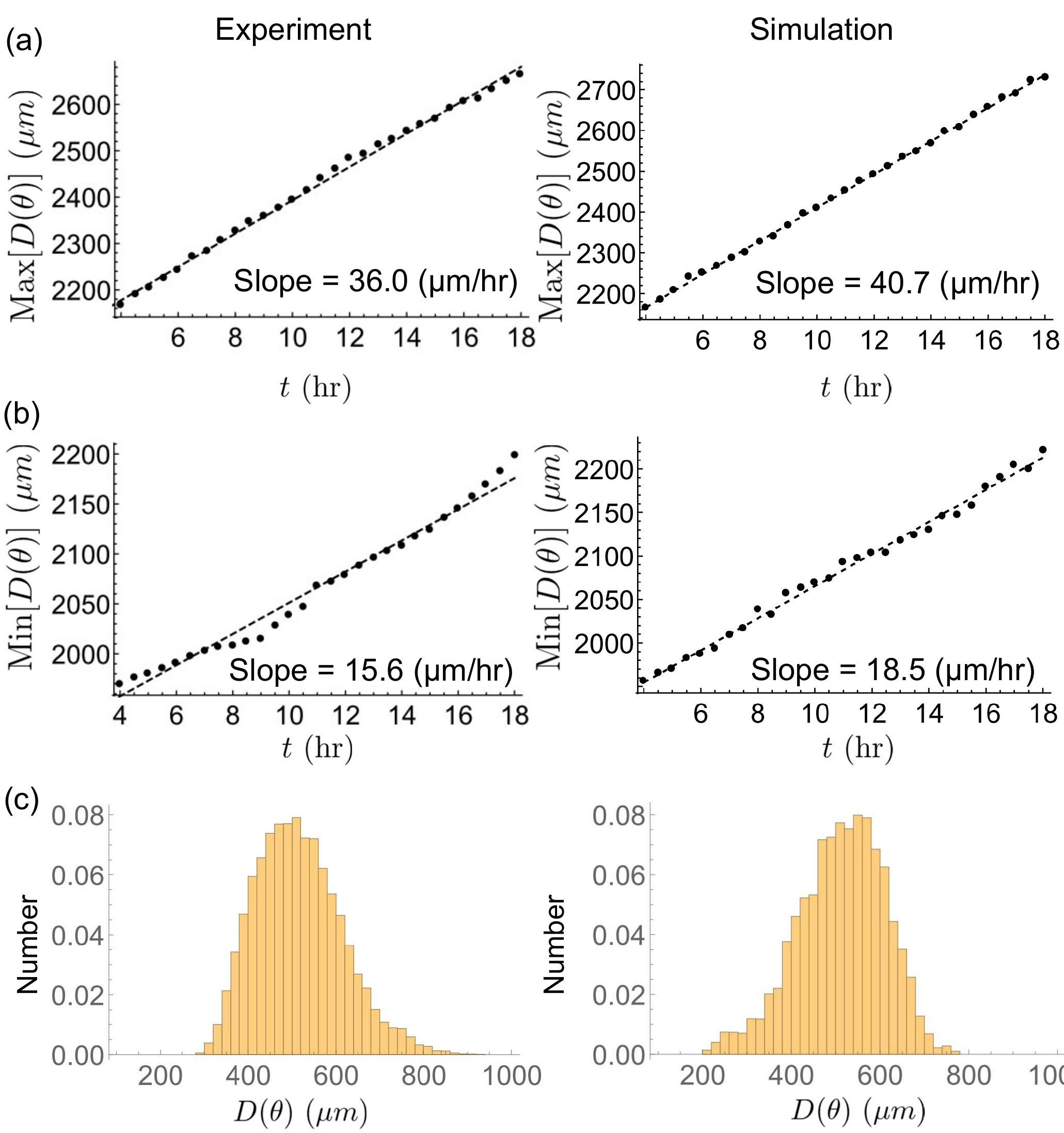}
\caption{\label{fig:10}Comparison of the experimentally and numerically obtained results with the collagen-coated dish, corresponding to Fig. 6.
(a) Time evolution of the maximum values of $D(\theta)$.  (b) Time evolution of the minimum values of $D(\theta)$. (c) Distribution of the increment of $D(\theta)$ in 18 hr. 
Left panels show the experimentally obtained data ($n=4$), and the right panels show the numerically obtained data.
Parameters for numerical simulation:  $v_l = 126 \ [\mu m\ h^{-1}]$, $v_f = 17.87 \ [\mu m \ h^{-1}]$, $p_l =0.1$, $\rho = 6.67 [h^{-1}]$, and $\sigma=28.8 \  [\mu m \ h^{-1/2}]$.}
\end{figure*}